\documentclass[
 prd,
 reprint,
 superscriptaddress,
 amsmath,amssymb,
 aps,
 floatfix,
]{revtex4-2}

\usepackage{makecell}
\usepackage[english]{babel}
\usepackage{xcolor}
\usepackage{xspace}
\usepackage{graphicx}
\usepackage{siunitx}
\usepackage{upgreek}
\usepackage{multirow}
\usepackage{hyperref}
\usepackage{cleveref}

\newcommand{\ricochet}{\textsc{Ricochet}\xspace}
\newcommand{\cevns}{CE$\nu$NS\xspace}
\newcommand{\uMUX}{$\mu$MUX\xspace}
\newcommand{\uPhirtHz}{$\upmu \Phi_0/\sqrt{\text{Hz}}$\xspace}

\begin{document}

\title{Characterization of Aluminum Microwave SQUID Multiplexers for \cevns Detection}

\author{\mbox{J Amidei}}
\affiliation{Physics Department, Colorado School of Mines, Golden, Colorado 80401, USA}

\author{\mbox{A Armatol}}
\affiliation{Université Lyon 1, CNRS, IP2I, UMR 5822, Villeurbanne, France}

\author{\mbox{C Augier}}
\affiliation{Université Lyon 1, CNRS, IP2I, UMR 5822, Villeurbanne, France}

\author{\mbox{L Bailly-Salins}}
\affiliation{Université Grenoble Alpes, CNRS, Grenoble INP, LPSC-IN2P3, 38000 Grenoble, France}

\author{\mbox{G Baulieu}}
\affiliation{Université Lyon 1, CNRS, IP2I, UMR 5822, Villeurbanne, France}

\author{\mbox{L Bergé}}
\affiliation{Université Paris-Saclay, CNRS/IN2P3, IJCLab, 91405 Orsay, France}

\author{\mbox{J Billard}}
\affiliation{Université Lyon 1, CNRS, IP2I, UMR 5822, Villeurbanne, France}

\author{\mbox{J Blé}}
\affiliation{Université Grenoble Alpes, CNRS, Grenoble INP, LPSC-IN2P3, 38000 Grenoble, France}

\author{\mbox{G Brenot}}
\affiliation{Université Grenoble Alpes, CNRS, Grenoble INP, LPSC-IN2P3, 38000 Grenoble, France}

\author{\mbox{G Bres}}
\affiliation{Université Grenoble Alpes, CNRS, Grenoble INP, Institut Néel, 38000 Grenoble, France}

\author{\mbox{J-L Bret}}
\affiliation{Université Grenoble Alpes, CNRS, Grenoble INP, Institut Néel, 38000 Grenoble, France}

\author{\mbox{A Broniatowski}}
\affiliation{Université Paris-Saclay, CNRS/IN2P3, IJCLab, 91405 Orsay, France}

\author{\mbox{M Calvo}}
\affiliation{Université Grenoble Alpes, CNRS, Grenoble INP, Institut Néel, 38000 Grenoble, France}

\author{\mbox{A Cavanna}}
\affiliation{Université Paris-Saclay, CNRS, C2N, Palaiseau, 91120 Palaiseau, France}

\author{\mbox{A Cazes}}
\affiliation{Université Lyon 1, CNRS, IP2I, UMR 5822, Villeurbanne, France}

\author{\mbox{E Celi}}
\affiliation{Department of Physics and Astronomy, Northwestern University, Evanston, Illinois 60208, USA}

\author{\mbox{D Chaize}}
\affiliation{Université Lyon 1, CNRS, IP2I, UMR 5822, Villeurbanne, France}

\author{\mbox{M Chala}}
\affiliation{Université Grenoble Alpes, CNRS, Grenoble INP, LPSC-IN2P3, 38000 Grenoble, France}

\author{\mbox{M Chapellier}}
\affiliation{Université Paris-Saclay, CNRS/IN2P3, IJCLab, 91405 Orsay, France}

\author{\mbox{L Chaplinsky}}
\affiliation{Department of Physics, University of Massachusetts at Amherst, Amherst, Massachusetts 01003, USA}

\author{\mbox{R Chen}}
\affiliation{Department of Physics and Astronomy, Northwestern University, Evanston, Illinois 60208, USA}

\author{\mbox{I Cojocari}}
\affiliation{Université Paris-Saclay, CNRS/IN2P3, IJCLab, 91405 Orsay, France}

\author{\mbox{J Colas}}
\affiliation{Université Lyon 1, CNRS, IP2I, UMR 5822, Villeurbanne, France}

\author{\mbox{L Couraud}}
\affiliation{Université Paris-Saclay, CNRS, C2N, Palaiseau, 91120 Palaiseau, France}

\author{\mbox{E Cudmore}}
\affiliation{Department of Physics, University of Toronto, Toronto, Ontario M5S 1A7, Canada}

\author{\mbox{M De Jesus}}
\affiliation{Université Lyon 1, CNRS, IP2I, UMR 5822, Villeurbanne, France}

\author{\mbox{P de Marcillac}}
\affiliation{Université Paris-Saclay, CNRS/IN2P3, IJCLab, 91405 Orsay, France}

\author{\mbox{N Dombrowski}}
\affiliation{Laboratory for Nuclear Science, Massachusetts Institute of Technology, Cambridge, Massachusetts 02139, USA}

\author{\mbox{L Dumoulin}}
\affiliation{Université Paris-Saclay, CNRS/IN2P3, IJCLab, 91405 Orsay, France}

\author{\mbox{A Durnez}}
\affiliation{Université Paris-Saclay, CNRS, C2N, Palaiseau, 91120 Palaiseau, France}

\author{\mbox{R Faure}}
\affiliation{Université Lyon 1, CNRS, IP2I, UMR 5822, Villeurbanne, France}

\author{\mbox{S Ferriol}}
\affiliation{Université Lyon 1, CNRS, IP2I, UMR 5822, Villeurbanne, France}

\author{\mbox{E Figueroa-Feliciano}}
\affiliation{Department of Physics and Astronomy, Northwestern University, Evanston, Illinois 60208, USA}

\author{\mbox{J A Formaggio}}
\affiliation{Laboratory for Nuclear Science, Massachusetts Institute of Technology, Cambridge, Massachusetts 02139, USA}
\affiliation{Department of Physics, Massachusetts Institute of Technology, Cambridge, Massachusetts 02139, USA}

\author{\mbox{S Fuard}}
\affiliation{Institut Laue-Langevin, CS 20156, 38042 Grenoble Cedex 9, France}

\author{\mbox{J Gascon}}
\affiliation{Université Lyon 1, CNRS, IP2I, UMR 5822, Villeurbanne, France}

\author{\mbox{C Girard-Carillo}}
\affiliation{Université Grenoble Alpes, CNRS, Grenoble INP, LPSC-IN2P3, 38000 Grenoble, France}

\author{\mbox{A Giuliani}}
\affiliation{Université Paris-Saclay, CNRS/IN2P3, IJCLab, 91405 Orsay, France}

\author{\mbox{C Goy}}
\affiliation{Université Grenoble Alpes, CNRS, Grenoble INP, LPSC-IN2P3, 38000 Grenoble, France}

\author{\mbox{C Guerin}}
\affiliation{Université Lyon 1, CNRS, IP2I, UMR 5822, Villeurbanne, France}

\author{\mbox{L Haegel}}
\affiliation{Université Lyon 1, CNRS, IP2I, UMR 5822, Villeurbanne, France}

\author{\mbox{P M Harrington}}
\affiliation{Laboratory for Nuclear Science, Massachusetts Institute of Technology, Cambridge, Massachusetts 02139, USA}
\affiliation{Research Laboratory of Electronics, Massachusetts Institute of Technology, Cambridge, Massachusetts 02139, USA}

\author{\mbox{S A Hertel}}
\affiliation{Department of Physics, University of Massachusetts at Amherst, Amherst, Massachusetts 01003, USA}

\author{\mbox{C F Hirjibehedin}}
\affiliation{Lincoln Laboratory, Massachusetts Institute of Technology, Lexington, Massachusetts 02421, USA}

\author{\mbox{C Hoarau}}
\affiliation{Université Grenoble Alpes, CNRS, Grenoble INP, LPSC-IN2P3, 38000 Grenoble, France}

\author{\mbox{Z Hong}}
\affiliation{Department of Physics, University of Toronto, Toronto, Ontario M5S 1A7, Canada}

\author{\mbox{D Howard}}
\affiliation{Physics Department, Colorado School of Mines, Golden, Colorado 80401, USA}

\author{\mbox{J-C Ianigro}}
\affiliation{Université Lyon 1, CNRS, IP2I, UMR 5822, Villeurbanne, France}

\author{\mbox{Y Jin}}
\affiliation{Université Paris-Saclay, CNRS, C2N, Palaiseau, 91120 Palaiseau, France}

\author{\mbox{A Juillard}}
\affiliation{Université Lyon 1, CNRS, IP2I, UMR 5822, Villeurbanne, France}

\author{\mbox{D Karaivanov}}
\affiliation{Department of Nuclear Spectroscopy and Radiochemistry, Laboratory of Nuclear Problems, JINR, Dubna, Moscow Region 141980, Russia}

\author{\mbox{T Khussainov}}
\altaffiliation{also at: Institute of Nuclear Physics of the Ministry of Energy of the Republic of Kazakhstan, 1 Ibragimov Street, 050032, Almaty, Kazakhstan}
\affiliation{Department of Nuclear Spectroscopy and Radiochemistry, Laboratory of Nuclear Problems, JINR, Dubna, Moscow Region 141980, Russia}

\author{\mbox{A Kubik}}
\altaffiliation{also at: SNOLAB, Creighton Mine \#9, 1039 Regional Road 24, Sudbury, ON P3Y 1N2, Canada}
\affiliation{Department of Physics, University of Toronto, Toronto, Ontario M5S 1A7, Canada}

\author{\mbox{J Lamblin}}
\affiliation{Université Grenoble Alpes, CNRS, Grenoble INP, LPSC-IN2P3, 38000 Grenoble, France}

\author{\mbox{T Le Bellec}}
\affiliation{Université Lyon 1, CNRS, IP2I, UMR 5822, Villeurbanne, France}

\author{\mbox{L Leroy}}
\affiliation{Université Paris-Saclay, CNRS, C2N, Palaiseau, 91120 Palaiseau, France}

\author{\mbox{M Li}}
\affiliation{Laboratory for Nuclear Science, Massachusetts Institute of Technology, Cambridge, Massachusetts 02139, USA}
\affiliation{Department of Physics, Massachusetts Institute of Technology, Cambridge, Massachusetts 02139, USA}

\author{\mbox{A Lubashevskiy}}
\affiliation{Department of Nuclear Spectroscopy and Radiochemistry, Laboratory of Nuclear Problems, JINR, Dubna, Moscow Region 141980, Russia}

\author{\mbox{S Marnieros}}
\affiliation{Université Paris-Saclay, CNRS/IN2P3, IJCLab, 91405 Orsay, France}

\author{\mbox{R Martin}}
\affiliation{Université Grenoble Alpes, CNRS, Grenoble INP, LPSC-IN2P3, 38000 Grenoble, France}

\author{\mbox{N Martini}}
\affiliation{Université Lyon 1, CNRS, IP2I, UMR 5822, Villeurbanne, France}

\author{\mbox{J Minet}}
\affiliation{Université Grenoble Alpes, CNRS, Grenoble INP, Institut Néel, 38000 Grenoble, France}

\author{\mbox{A Monfardini}}
\affiliation{Université Grenoble Alpes, CNRS, Grenoble INP, Institut Néel, 38000 Grenoble, France}

\author{\mbox{F Mounier}}
\affiliation{Université Lyon 1, CNRS, IP2I, UMR 5822, Villeurbanne, France}

\author{\mbox{B M Niedzielski}}
\affiliation{Lincoln Laboratory, Massachusetts Institute of Technology, Lexington, Massachusetts 02421, USA}

\author{\mbox{V Novati}}
\affiliation{Université Grenoble Alpes, CNRS, Grenoble INP, LPSC-IN2P3, 38000 Grenoble, France}

\author{\mbox{W D Oliver}}
\affiliation{Department of Physics, Massachusetts Institute of Technology, Cambridge, Massachusetts 02139, USA}
\affiliation{Research Laboratory of Electronics, Massachusetts Institute of Technology, Cambridge, Massachusetts 02139, USA}
\affiliation{Electrical Engineering and Computer Science Department, Massachusetts Institute of Technology, Cambridge, Massachusetts 02139, USA}

\author{\mbox{E Olivieri}}
\affiliation{Université Paris-Saclay, CNRS/IN2P3, IJCLab, 91405 Orsay, France}

\author{\mbox{H D Pinckney}}
\affiliation{Laboratory for Nuclear Science, Massachusetts Institute of Technology, Cambridge, Massachusetts 02139, USA}
\affiliation{Research Laboratory of Electronics, Massachusetts Institute of Technology, Cambridge, Massachusetts 02139, USA}

\author{\mbox{D V Poda}}
\affiliation{Université Paris-Saclay, CNRS/IN2P3, IJCLab, 91405 Orsay, France}

\author{\mbox{D Ponomarev}}
\altaffiliation{also at: Lebedev Physical Institute of the Russian Academy of Sciences, 53 Leninskiy Prospect, 119991, Moscow, Russia}
\affiliation{Department of Nuclear Spectroscopy and Radiochemistry, Laboratory of Nuclear Problems, JINR, Dubna, Moscow Region 141980, Russia}

\author{\mbox{J-S Real}}
\affiliation{Université Grenoble Alpes, CNRS, Grenoble INP, LPSC-IN2P3, 38000 Grenoble, France}

\author{\mbox{F C Reyes}}
\affiliation{Laboratory for Nuclear Science, Massachusetts Institute of Technology, Cambridge, Massachusetts 02139, USA}
\affiliation{Department of Physics, Massachusetts Institute of Technology, Cambridge, Massachusetts 02139, USA}

\author{\mbox{A Rodriguez}}
\affiliation{Department of Physics and Astronomy, Northwestern University, Evanston, Illinois 60208, USA}

\author{\mbox{S Rozov}}
\affiliation{Department of Nuclear Spectroscopy and Radiochemistry, Laboratory of Nuclear Problems, JINR, Dubna, Moscow Region 141980, Russia}

\author{\mbox{I Rozova}}
\affiliation{Department of Nuclear Spectroscopy and Radiochemistry, Laboratory of Nuclear Problems, JINR, Dubna, Moscow Region 141980, Russia}

\author{\mbox{B Ryan}}
\affiliation{Laboratory for Nuclear Science, Massachusetts Institute of Technology, Cambridge, Massachusetts 02139, USA}

\author{\mbox{D Sabhari}}
\affiliation{Department of Physics and Astronomy, Northwestern University, Evanston, Illinois 60208, USA}

\author{\mbox{S Scorza}}
\affiliation{Université Grenoble Alpes, CNRS, Grenoble INP, LPSC-IN2P3, 38000 Grenoble, France}

\author{\mbox{K Serniak}}
\affiliation{Research Laboratory of Electronics, Massachusetts Institute of Technology, Cambridge, Massachusetts 02139, USA}
\affiliation{Lincoln Laboratory, Massachusetts Institute of Technology, Lexington, Massachusetts 02421, USA}

\author{\mbox{R Serra}}
\affiliation{Université Grenoble Alpes, CNRS, Grenoble INP, LPSC-IN2P3, 38000 Grenoble, France}
\affiliation{Institut Laue-Langevin, CS 20156, 38042 Grenoble Cedex 9, France}

\author{\mbox{Ye Shevchik}}
\affiliation{Department of Nuclear Spectroscopy and Radiochemistry, Laboratory of Nuclear Problems, JINR, Dubna, Moscow Region 141980, Russia}

\author{\mbox{T Soldner}}
\affiliation{Institut Laue-Langevin, CS 20156, 38042 Grenoble Cedex 9, France}

\author{\mbox{H Stickler}}
\affiliation{Lincoln Laboratory, Massachusetts Institute of Technology, Lexington, Massachusetts 02421, USA}

\author{\mbox{C Stone-Whitehead}}
\affiliation{Physics Department, Colorado School of Mines, Golden, Colorado 80401, USA}

\author{\mbox{Ch Ulysse}}
\affiliation{Université Paris-Saclay, CNRS, C2N, Palaiseau, 91120 Palaiseau, France}

\author{\mbox{W Van De Pontseele}}
\email{wouter.vandepontseele@mines.edu}
\affiliation{Physics Department, Colorado School of Mines, Golden, Colorado 80401, USA}

\author{\mbox{S Vasilyev}}
\affiliation{Department of Nuclear Spectroscopy and Radiochemistry, Laboratory of Nuclear Problems, JINR, Dubna, Moscow Region 141980, Russia}

\author{\mbox{P Vittaz}}
\affiliation{Université Lyon 1, CNRS, IP2I, UMR 5822, Villeurbanne, France}

\author{\mbox{S Weber}}
\affiliation{Lincoln Laboratory, Massachusetts Institute of Technology, Lexington, Massachusetts 02421, USA}

\author{\mbox{E Wolfe}}
\affiliation{Physics Department, Colorado School of Mines, Golden, Colorado 80401, USA}

\author{\mbox{W Woods}}
\affiliation{Lincoln Laboratory, Massachusetts Institute of Technology, Lexington, Massachusetts 02421, USA}

\author{\mbox{E Yakushev}}
\affiliation{Department of Nuclear Spectroscopy and Radiochemistry, Laboratory of Nuclear Problems, JINR, Dubna, Moscow Region 141980, Russia}

\author{\mbox{J Yang}}
\email{jiatongy@mit.edu}
\affiliation{Laboratory for Nuclear Science, Massachusetts Institute of Technology, Cambridge, Massachusetts 02139, USA}
\affiliation{Department of Physics, Massachusetts Institute of Technology, Cambridge, Massachusetts 02139, USA}

\author{\mbox{D Zinatulina}}
\affiliation{Department of Nuclear Spectroscopy and Radiochemistry, Laboratory of Nuclear Problems, JINR, Dubna, Moscow Region 141980, Russia}

\date{\today}

\begin{abstract}
We present the design, fabrication, and characterization of an aluminum-based six-channel microwave SQUID multiplexer (\uMUX) prototype for transition-edge sensor (TES) readout in the \ricochet experiment. The device consists of aluminum coplanar-waveguide resonators and RF SQUIDs with Dolan-style Al/AlO$_x$/Al Josephson junctions. By measuring the resonator scattering parameters at a range of probe tone frequencies, powers, and flux bias points, we demonstrate agreement between the device response and existing multiplexer models. We also characterize the noise performance in both open-loop and flux-ramping modes. With a high electron mobility transistor (HEMT) amplifier, open-loop measurements yield a flux sensitivity of 1--1.5~\uPhirtHz. With flux-ramp modulation, low-frequency $1/f$ noise is suppressed, and the flux sensitivity is around 3--4~\uPhirtHz, corresponding to a current sensitivity of 24--33~$\mathrm{pA}/\sqrt{\mathrm{Hz}}$ at the input coil. We further demonstrate a reduction in readout noise by incorporating a Josephson traveling-wave parametric amplifier (JTWPA) between the \uMUX and the HEMT. This achieves an open-loop flux sensitivity of 0.3--0.6~\uPhirtHz and an effective system noise temperature below \qty{1}{\K}. These results establish aluminum $\mu$MUX devices as a viable and extensible readout technology for low-noise cryogenic detector arrays.


\end{abstract}

\maketitle

%-----------------------------------------------
\section{Introduction}

In recent years, there has been growing interest among particle physics experiments in detectors with lower energy thresholds and improved energy resolution. Examples of these experiments include searches for dark matter~\cite{armengaud_constraints_2016, agnese_first_2018, angloher_results_2023, alloni_new_2025}, coherent elastic neutrino--nucleus scattering (\cevns)~\cite{angloher_exploring_2019, augier_ricochet_2023}, neutrinoless double beta decay~\cite{legend_design_2021, the_cuore_collaboration_search_2022, agrawal_development_2025, cupidcollaboration_cupid_2025}, and measurements of the neutrino mass~\cite{gastaldo_electron_2014, faverzani_holmes_2016}. For thermal detectors, the energy resolution scales with the square root of the target heat capacity~\cite{ascheron_thermal_2005}, so a natural strategy is to operate them at cryogenic temperatures and to read them out with low-temperature sensors such as transition-edge sensors (TESs)~\cite{irwin_application_1995, ascheron_transition-edge_2005}, metallic magnetic calorimeters (MMCs)~\cite{enss_metallic_2000, kempf_physics_2018}, germanium neutron transmutation doped thermistors~\cite{haller_advanced_1994}, and kinetic inductance detectors~\cite{day_broadband_2003, zmuidzinas_superconducting_2012}. The exposure scales with target mass. On the other hand, the energy resolution depends on properties such as heat capacity and the propagation length of energy carriers (e.g. phonons, electron-hole pairs, or Bogoliubov quasiparticles~\cite{catelani_using_2022}), which constrain the size of each pixel. To accommodate both event rate and energy resolution constraints, many experiments tend to construct detector arrays.

One such experiment is \ricochet, which aims to detect reactor \cevns events at the Institut Laue--Langevin research reactor in Grenoble, France. \ricochet has two cryogenic detector arrays with distinct target materials. The CryoCube detector uses germanium crystals with both thermal and ionization readout channels for discrimination between nuclear and electron recoils. The recently deployed modules at the reactor achieved baseline ionization resolution of 40 eVee and baseline phonon resolution of 50--80 eV~\cite{armatol_characterization_2025}. The Q-Array detector uses superconducting targets, such as tin crystals, coupled to TES sensors~\cite{augier_results_2023}. Particle-induced phonons and quasiparticles in the targets are converted into current signals at the TES. However, reading out large arrays of TESs is challenging as the wiring complexity, cryogenic heat load, and total system cost all scales linearly with the number of individually instrumented channels. This motivates the use of microwave superconducting quantum interference device (SQUID) multiplexers (\uMUX), a widely used solution for scaling up low impedance current sensors such as TESs~\cite{bennett_integration_2015,mates_simultaneous_2017,becker_working_2019,szypryt_tabletop_2023,satterthwaite_simons_2024} and MMCs~\cite{kempf_design_2017, neidig_full-scale_2025}.

Over the past few decades, substantial effort has been devoted to the fabrication and modeling of \uMUX devices. While all \uMUX implementations to date used niobium~\cite{mates_microwave_2011, kempf_design_2017, nakashima_adjustable_2017, song_design_2026}, aluminum has also emerged as a mature and widely used material platform for superconducting microwave devices. Aluminum benefits from a well-controlled native oxide that enables simple and reproducible Josephson junction fabrication through shadow evaporation. This approach has supported the realization of high internal quality factor resonators and Josephson junctions~\cite{melville_comparison_2020, gingras_improving_2025}. Motivated by these considerations, we report our progress toward developing an aluminum-based \uMUX suitable for the TES readout of \ricochet.

We present the design, fabrication, and characterization of a six-channel aluminum \uMUX. We first introduce the principles of microwave SQUID multiplexing and describe the model used to capture the device response in \cref{sec:model}. In \cref{sec:design_fab}, we discuss the design choices and outline the fabrication process. The experimental setup used for device characterization is described in \cref{sec:setup}. Measurement results are presented in \cref{sec:results}, where we first measure the transmission coefficient, $S_{21}$, of the \uMUX at different operating points using a vector network analyzer (VNA) and compare the results to existing models. We then characterize the noise of the \uMUX channels with a high electron mobility transistor (HEMT) amplifier and find performance comparable to existing niobium devices. Finally, we explore approaches to lowering the noise by adding a Josephson traveling-wave parametric amplifier (JTWPA) to the readout chain, which demonstrates improved noise performance under both open-loop and flux-ramp operations.

\begin{figure}[htb]
    \centering
    \includegraphics[width=\linewidth]{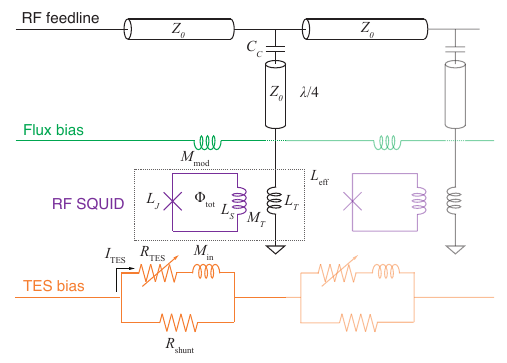}
    \caption{Schematic of the microwave SQUID multiplexer (\uMUX) showing two channels. Each channel consists of a quarter-wave ($\lambda/4$) resonator capacitively coupled to a common RF feedline with characteristic impedance $Z_0$ via coupling capacitor $C_c$. The effective inductance of the resonator termination is $L_{\mathrm{eff}}$, which arises from the self-inductance of the resonator termination $L_T$, an RF SQUID with a loop inductance $L_S$, and a Josephson junction inductance $L_J$. The SQUID is inductively coupled to the resonator with mutual inductance $M_T$, to a common flux bias line with mutual inductance $M_{\mathrm{mod}}$, and to the TES input coil with mutual inductance $M_{\mathrm{in}}$. The total flux through the SQUID loop is $\Phi_{\mathrm{tot}}$. The TES circuit consists of a TES with resistance $R_{\mathrm{TES}}$, shunt resistor $R_{\mathrm{shunt}}$, and a bias line providing the bias current for all TESs. The TES current $I_{\mathrm{TES}}$ is inductively coupled to the SQUID, enabling readout via shifts in the resonator response.}

    \label{fig:multiplexer_schematic}
\end{figure}

%-----------------------------------------------
\section{Multiplexer model}
\label{sec:model}

\subsection{\texorpdfstring{\uMUX model}{muMUX model}}
\label{model:uMUX_model}

Figure \ref{fig:multiplexer_schematic} shows a schematic of the \uMUX coupled to an array of TESs. Each channel of the multiplexer consists of an RF SQUID coupled inductively to a quarter-wave resonator. The resonators, properly spaced in frequency, are capacitively coupled to a common RF feedline in a hanger geometry. The working principle of this type of $\mu$MUX is detailed in~\cite{mates_microwave_2011}. The RF SQUID is inductively coupled to the TES circuit through a mutual inductance $M_{\text{in}}$. Thus a change in TES current results in a change in the total magnetic flux through the RF SQUID loop. This current changes the effective inductance $L_{\text{eff}}$ of the resonator termination (grey box in figure \ref{fig:multiplexer_schematic}), resulting in a shift in the resonant frequency, which is detected as a change in the feedline's transmission coefficient, $S_{21}$, near resonance. Thus each TES current signal $I_{\text{TES}}(t)$ can be reconstructed from each resonance's $S_{21}(t)$, which is probed with a single RF feedline.

The quarter-wave resonator has a bare resonant frequency $f_0 = c_0 / (4 l_r)$, where $c_0$ is the effective phase velocity in the coplanar waveguide and $l_r$ is the resonator length. When the substrate thickness is much larger than the gap between the center conductor and the ground plane, the effective dielectric constant of the coplanar waveguide (CPW) is $\epsilon_{\mathrm{CPW}} = (\epsilon_s + 1)/2$~\cite{steer_microwave_2019}, where $\epsilon_s$ is the dielectric constant of the substrate. For a silicon substrate, this yields $c_0 = c / \sqrt{\epsilon_{\mathrm{CPW}}} \approx 0.4c$, where $c$ is the speed of light in vacuum. For a quarter-wave resonator capacitively coupled to the RF feedline with capacitance $C_c$ and inductively loaded by $L_{\text{eff}}$, the resonant frequency is shifted to~\cite{kempf_design_2017}
\begin{gather} \label{eq:f_res_vs_L_T}
    f_r \approx f_0 - 4 f_0^2 \left( C_c Z_0 + \frac{L_{\text{eff}}}{Z_0} \right)
\end{gather}
assuming $|f_r - f_0| \ll f_0$. Here, $Z_0$ is the characteristic impedance of the coaxial cable and the resonator, typically around \qty{50}{\ohm}.

To understand the resonator's response and sensitivity to the TES current, we need to consider the effect of the junction's nonlinear inductance. The total flux threading the RF SQUID satisfies the relation~\cite{wegner_analytical_2022}
\begin{equation} \label{eq:tot_flux}
\Phi_{\mathrm{tot}} = \Phi_{\mathrm{ext}} + \Phi_{\mathrm{rf}} \sin(\omega t) + \frac{\Phi_0 L_S}{2\pi L_J} \sin\left( \frac{2\pi \Phi_{\mathrm{tot}}}{\Phi_0} \right),
\end{equation}
where $\Phi_{\mathrm{ext}}$ is the applied flux, the $\Phi_{\mathrm{rf}}$ term is the AC flux coupled in from the resonator, and the last term is the self-screening flux from the SQUID loop. Here $L_S$ is the self-inductance of the SQUID loop, and we define the Josephson inductance as
\begin{equation} \label{eq:L_J_def}
    L_J\equiv\Phi_0/2\pi I_c,
\end{equation}
where $I_c$ is the critical current of the junction, and $\Phi_0$ is the magnetic flux quantum. Note that equation~\eqref{eq:tot_flux} is implicit, so $\Phi_{\text{tot}}$ has to be solved self-consistently with given values of $\Phi_{\text{ext}}$ and $\Phi_{\text{rf}}$. We operate in the $\beta_L\equiv L_S/L_J<1$ regime, ensuring the mapping from $\Phi_{\text{tot}}$ to $\Phi_{\text{ext}}$ remains single-valued. When operating with probe tone frequency, $f_{\text{exc}}$, within the resonance linewidth, the relation between the RF flux amplitude $\Phi_{\text{rf}}$ and probe tone power $P_{\text{exc}}$ can be approximated as~\cite{wegner_analytical_2022}
\begin{gather} \label{eq:Phi_rf}
    \Phi_{\text{rf}} = M_T I_T \approx M_T \sqrt{\frac{16 Q_l^2 P_{\text{exc}}}{\pi Q_c Z_0}}.
\end{gather}
Here $M_T$ is the mutual inductance between the SQUID loop and the resonator termination, $I_T$ is the amplitude of the AC current in the termination of the resonator, and $Q_l$ and $Q_c$ are the loaded and coupled quality factors of the resonator.

In the low probe-power limit ($\Phi_{\mathrm{rf}} \ll \Phi_0$), the AC $\Phi_{\text{rf}}$ term in equation \eqref{eq:tot_flux} can be ignored. Thus the SQUID modifies the resonator termination inductance as \cite{mates_microwave_2011}
\begin{gather} \label{eq:L_eff_low_power}
    L_{\text{eff}} = L_T - \frac{M_T^2}{L_S} \frac{\beta_L \cos(\varphi_{\text{tot}})}{1+\beta_L \cos(\varphi_{\text{tot}})},
\end{gather}
where $L_T$ is the self-inductance of the resonator termination. We also define the contributions to the phase difference of the superconducting wavefunction across the junction $\varphi_k = 2\pi \Phi_k/\Phi_0$, where \mbox{$k\in\{\text{rf}, \text{ext}, \text{tot}\}$}. Then we can substitute equation \eqref{eq:L_eff_low_power} into equation \eqref{eq:f_res_vs_L_T} and get
\begin{gather} \label{eq:squid_dispersion_low_pwr}
f_{\text{r}} \approx f_0 - 4f_0^2\left[C_c Z_0 + \frac{L_T}{Z_0} - \frac{M_T^2}{Z_0 L_S} \frac{\beta_L \cos{\varphi_{\text{tot}}}}{1+\beta_L \cos{\varphi_{\text{tot}}}}\right].
\end{gather}
This models how the resonant frequency shifts with external flux through the SQUID.

At higher probe powers, the AC flux modifies the Josephson response. It is important to model this behavior since the operating point with optimal flux sensitivity often lies in this regime. This effect can be captured by a series expansion,
\begin{equation} \label{eq:L_eff_finite_power}
L_{\mathrm{eff}} \approx L_T - \frac{M_T^2}{L_S}
\frac{2\beta_L}{\varphi_{\mathrm{rf}}}
\sum_{m,n} p_{m,n},
\end{equation}
where $p_{m,n} = a_{m,n} \beta_L^{b_{m,n}} J_1(c_{m,n}\varphi_{\text{rf}}) \cos(c_{m,n} \varphi_{\text{ext}})$. Here $J_1$ is the Bessel function of the first kind, $n$ and $m$ are indices of a series expansion, and the values of the coefficients $a, b,$ and $c$ are documented in table 1 of \cite{wegner_analytical_2022}. Combining equations \eqref{eq:L_eff_finite_power} and \eqref{eq:f_res_vs_L_T}, the resonant frequency $f_r$ with probe tone power $P_{\text{exc}}$ at a flux bias point $\Phi_{\text{ext}}$ is
\begin{gather} \label{eq:squid_dispersion_finite_pwr}
    f_r \approx f_0 - 4 f_0^2 \left[ C_c Z_0 + \frac{L_T}{Z_0} - \frac{M_T^2}{Z_0 L_S} \frac{2 \beta_L}{\varphi_{\text{rf}}} \sum_{m,n} p_{m,n}\right].
\end{gather}
Following this expression, the peak-to-peak range of the $f_r$ shift, $\Delta f_{\text{pp}}$, decreases as $P_{\text{exc}}$ increases. This is effectively a result of an increased AC flux amplitude causing the resonance shift over different flux bias points to average out.

To map $f_r$ to the measured voltage at the digitizer, we need the transmission coefficient, $S_{21}$, of the RF line through the multiplexer. This is given by
\begin{gather} \label{eq:S_21}
    S_{21} = 1- \frac{(Q_l/Q_c)}{1+2iQ_l(f_{\text{exc}}/f_r-1)},
\end{gather}
where $i$ is the imaginary unit. The quality factors follow the expression $Q_l^{-1} = Q_i^{-1} + {Q_c}^{-1}$, where $Q_i$ is the resonator's internal quality factor. The $S_{21}$ values of the resonator at different probe tone frequencies form a circle in the complex plane. It is centered at $S_{21} = 1-Q_l/2Q_c$ and intersects with the real axis at 1. In this paper, we read out the time series of $S_{21}$ on resonance in both real and imaginary quadratures.

\subsection{Flux-ramp modulation}
\label{model:FRM}
The flux-ramp modulation is the key that enables the readout of all channels using a common flux bias line, which is inductively coupled to each SQUID with mutual inductance $M_\text{mod}$~\cite{mates_flux-ramp_2012}. In FRM, a sawtooth flux signal with amplitude $n_{\text{FRM}}$ flux quanta and frequency $f_{\mathrm{ramp}}$ is applied to bias the SQUIDs resulting in the $S_{21}$ signal modulated at a frequency \mbox{$f_{\mathrm{mod}} = n_{\text{FRM}} f_{\mathrm{ramp}}$}.

We record the time series $x(t)$, where $x$ may correspond to the signal readout in whichever domain, such as $\mathrm{Im}(S_{21})$ or $|S_{21}|$. The ramp frequency $f_{\mathrm{ramp}}$ is chosen to be much higher than the bandwidth of the TES-induced flux signal, with an upper limit around the resonance linewidth~\cite{yu_bandwidth_2022}. Thus the TES signal appears as a quasi-static flux offset within each flux-ramp period, resulting in a phase shift $\Delta \phi$ relative to the signal without the TES-induced flux. When $x(t)$ is roughly sinusoidal, the phase shift $\Delta\phi$ within a flux-ramp interval can be extracted from $x(t)$ using
\begin{gather}
    \Delta\phi = \arctan \left( \frac{\sum_i x(t_i) \sin(2\pi f_{\mathrm{mod}} t_i)}{\sum_i x(t_i) \cos(2\pi f_{\mathrm{mod}} t_i)} \right),
\end{gather}
where the sums of discrete time steps are taken over an integer number of modulation periods. The external flux can then be reconstructed using the linear relation $\Phi_{\mathrm{ext}} = \Phi_0 \Delta\phi / 2\pi$.

\subsection{Sensitivity estimation} \label{model:sensitivity}

The open-loop readout of the \uMUX is defined as measuring the multiplexer $S_{21}$ at a fixed flux bias point (no FRM)~\cite{schuster_simulation_2023, redondo_optimal_2024}. Two main noise contributions limit the flux sensitivity in this readout scheme: $1/f$ noise that dominates at low frequencies and white amplifier noise that dominates at high frequencies. In this section, we only model the white noise contribution to flux sensitivity. The full frequency dependence of the measured noise is presented in \cref{sec:results}.

The flux sensitivity can be calculated with the multiplexer model discussed in \cref{model:uMUX_model}. In the open-loop readout mode, the flux sensitivity can be optimized by tuning the operating point $(f_{\mathrm{exc}}, \Phi_{\mathrm{ext}}, P_{\mathrm{exc}})$. The noise spectral density (NSD) of the input voltage of the first-stage amplifier is modeled as $S_V = k_B T_{\text{sys}} Z_0$, where $k_B$ is the Boltzmann constant and $T_{\mathrm{sys}}$ is the system noise temperature. When reading out in the imaginary quadrature, the NSD of $\mathrm{Im}(S_{21})$ is
\begin{equation} \label{eq:S_im}
    S^{\mathrm{Im}} = k_B T_{\text{sys}} / P_{\mathrm{exc}}.
\end{equation}
The flux-equivalent NSD when reading out in $\mathrm{Im}(S_{21})$ is obtained via the chain rule
\begin{equation} \label{eq:OL_noise_model}
    S^\text{Im}_\Phi = S^{\text{Im}} \left|\frac{\partial \Phi_{\text{ext}}}{\partial f_r}  \frac{\partial f_r}{\partial \text{Im}(S_{21})} \right|^2,
\end{equation}
where the partial derivatives can be obtained from equations \eqref{eq:squid_dispersion_finite_pwr} and \eqref{eq:S_21} at each operating point. Under this simple model, we take $\sqrt{S^\text{Im}_\Phi}$ to be the open-loop flux sensitivity.

In FRM, the amplifier noise floor is degraded relative to that of the open-loop at the optimal flux bias point. The ratio between the FRM and open-loop flux sensitivities, $c_{\mathrm{deg}}\equiv \sqrt{S_{\Phi,\mathrm{FRM}}/S_{\Phi, \text{OL}}}$, satisfies $c_{\mathrm{deg}} \geq \sqrt{2}$~\cite{mates_flux-ramp_2012}. This is because the flux-ramp signal sweeps the SQUID through regions of both high and low responsivity, which effectively averages the transfer function over one flux quantum. FRM sensitivity is further degraded by the finite duty cycle of the ramp: the rapid reset of the sawtooth waveform introduces transient irregularities at the beginning of each ramp interval, occupying a fraction $\alpha$ of the ramp period. These intervals must be discarded, and the resulting flux sensitivity under FRM is $\sqrt{S_{\Phi,\mathrm{FRM}}} = \sqrt{S_{\Phi}}\,\frac{c_{\mathrm{deg}}}{\sqrt{\alpha}}$.

Although FRM increases the amplifier noise compared to open-loop operation, it mitigates most of the low-frequency noise downstream of the SQUID~\cite{kempf_design_2017, mates_flux-ramp_2012}. Since they are at lower frequencies than the flux-ramp signal and occur after the SQUID, they appear as DC offsets or amplitude scaling within individual ramp intervals, which are averaged out during demodulation.

%-----------------------------------------------
\section{\uMUX design and fabrication}
\label{sec:design_fab}

\begin{figure*}[htb]
    \centering
    \includegraphics[width=\textwidth]{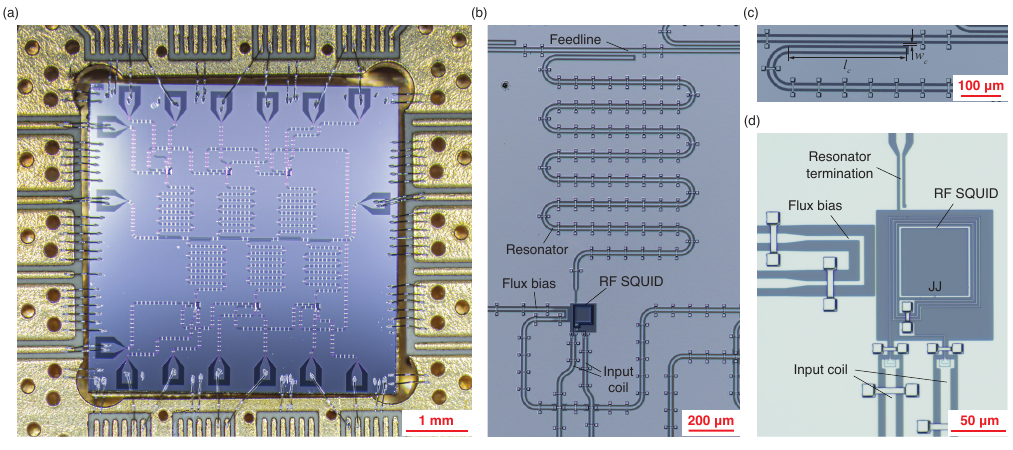}
    \caption{Photos of the multiplexer device. (a) Photo of the whole multiplexer chip. (b) Left: Zoom-in of one channel. (c) Zoom-in of the capacitive coupling region between the resonator and feedline. (d) Zoom-in of the RF SQUID region, showing how it is coupled to the flux bias line, input coil, and resonator.}
    \label{fig:device_photos}
\end{figure*}

The design of a multiplexer with six channels is shown in figure \ref{fig:device_photos}(a). We plan to increase the channel number up to $\mathcal{O}(100)$ per chip, which would allow us to scale up the \ricochet payload while reducing the cable complexity and heat load on the cryogenics. This device was fabricated at MIT Lincoln Laboratory on a 5$\times$\qty{5}{\mm^2} silicon substrate with high quality factor molecular beam epitaxial aluminum base metal~\cite{gingras_improving_2025}. The RF feedline and the resonators are CPW transmission lines (figure \ref{fig:device_photos}(b)). The center aluminum trace is \qty{9}{\um} wide. The distance between the center line and the ground plane is \qty{7}{\um}, resulting in a characteristic impedance of \mbox{$Z_0 =$ \qty{55}{\ohm}}. The aluminum air bridges short the ground planes on each side of the CPW, forcing them to be at the same potential and suppressing unwanted modes. The six resonators with varying lengths are spaced between 5 and \qty{5.5}{\GHz}. Resonators neighboring in frequency are placed away from each other to avoid crosstalk. Each resonator is capacitively coupled to the feedline via a parallel section of length $l_c$ and separation $w_c$ (see figure \ref{fig:device_photos}(c)). The coupled quality factor, $Q_c$, is determined by the length of this parallel coupler $l_c$ and the width of the ground plane separating the feedline and resonator $w_c$. In this design, we use $l_c=$ \qty{315}{\um} and $w_c=$ \qty{7}{\um}, targeting $Q_c=$ 25k at \qty{5.25}{\GHz}. We expect to have $Q_i \gg Q_c$ and thus $Q_l \approx Q_c$. Thus the estimated bandwidth of the resonator is $BW= (\pi f_r)/Q_l \approx$ \qty{660}{\kHz} at \qty{5.25}{\GHz}.

Each resonator is inductively coupled to an unshunted RF SQUID. The RF SQUID is a 63$\times$\qty{62}{\um^2} rectangle made out of a \qty{2}{\um} wide aluminum trace interrupted by an Al/AlOx/Al Josephson junction (see figure~\ref{fig:device_photos}(d)). The junction is fabricated using a Dolan bridge resist mask and double-angle shadow evaporation~\cite{gingras_improving_2025}. The designed critical current density of the junction is \qty{3}{\uA/\um^2}. The junction size is 1.45$\times$\qty{0.2}{\um^2}, which gives a designed junction inductance $L_J = $ \qty{378.3}{\pH}. The input coil has three turns around the SQUID. An air bridge closes the loop by connecting the innermost end of the input coil with the return path. The air bridge is necessary for this device geometry where the input coil is fabricated on one single layer. A common flux bias line passes by all SQUIDs without winding around them. COMSOL~\cite{comsol_multiphysics} simulations of the design yield the inductance matrix summarized in table~\ref{tab:COMSOL_sim}.

\begin{table}[htb]
    \centering
    \begin{tabular}{c c c c c}
    \hline
          &  \small{Input}& \small{Flux bias} & \small{SQUID} & \small{Resonator} \\
    \hline
    \small{Input} & 1800 & 5 & 267 & 35 \\
    \small{Flux bias} & -- & 152 & 1 & 1 \\
    \small{SQUID} & -- & -- & 191 & 6 \\
    \small{Resonator} & -- & -- & -- & 86 \\
    \hline
    \end{tabular}
    \caption{Inductance matrix of the circuit elements simulated with COMSOL. The following self-inductances are considered in the text: flux bias line self inductance ($L_{\mathrm{mod}}$), SQUID self inductance ($L_S$), and resonator self inductance ($L_T$).  The off-diagonal terms correspond to mutual inductance between: the SQUID and the resonator ($M_T$), the input coil and the SQUID ($M_{\mathrm{in}}$), the flux bias line and the SQUID ($M_{\mathrm{mod}}$), as well as the resonator and the flux bias line ($M_p$). The units are in pH.}
    \label{tab:COMSOL_sim}
\end{table}

%-----------------------------------------------
\section{Measurement setup}
\label{sec:setup}

The multiplexer is mounted on an extension panel attached to the mixing chamber of a Bluefors XLD dilution refrigerator, which reaches a base temperature of \qty{10}{mK}. The extension panel is enclosed in a multi-layer magnetic shield. From the innermost to the outermost layer, the shielding consists of aluminum, tin-plated copper, and two layers of Amumetal 4K. The RF lines connect to the K connectors on the multiplexer package. The flux bias line is a twisted pair with a \qty{391}{\kHz} low pass filter on the \qty{4}{\K} stage. The input coils remain electrically open during all measurements presented.

In sections \ref{results:low_rf_pwr} and \ref{results:power_calib}, we characterize the resonant frequencies, quality factors, and circuit parameters of the multiplexer. These results are obtained by measuring the $S_{21}$ of the RF chain with a Keysight E5080B VNA. Sections \ref{results:HEMT_noise} and \ref{results:TWPA} mainly present noise measurements, which are performed with a homodyne setup for better control of the sampling frequency.

\begin{figure}[htb]
    \centering
    \includegraphics[width=\columnwidth]{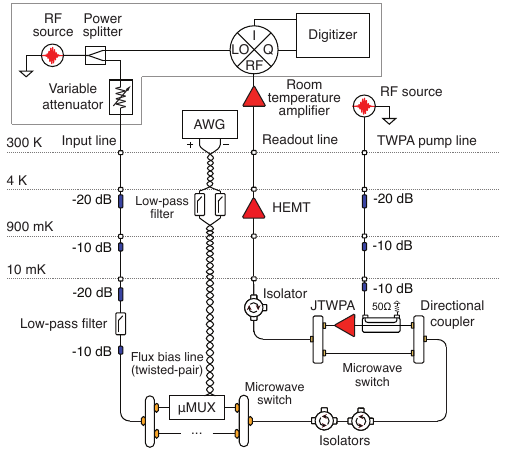}
    \caption{Schematic of the homodyne measurement setup.}
    \label{fig:setup_schematic}
\end{figure}

Figure \ref{fig:setup_schematic} shows a schematic of the homodyne measurement setup. When measuring with the VNA, everything in the gray box is replaced with the VNA. In the homodyne measurement setup, the signal from the RF source (BNC 865-M) is split into two branches with the power splitter. One branch serves as the local oscillator (LO) signal for the I/Q down-conversion. The other branch passes through a variable attenuator (Vaunix LDA-602), which is used to control the probe tone power, before entering the dilution refrigerator. Along this path, a total of \qty{60}{\dB} of attenuation is distributed across multiple temperature stages to suppress thermal noise. The multiplexer is installed between a pair of microwave switches (Radiall R591762600) in parallel with other devices in the fridge. The JTWPA is located between a second pair of switches (Teledyne CCR-33S60-N). The alternate branch of this switch has a through-line to bypass the JTWPA for measurements with the HEMT amplifier only. To prevent the JTWPA pump signal from entering the multiplexer, two isolators are placed between the JTWPA switch and the multiplexer. The JTWPA pump tone is attenuated by \qty{40}{\dB} before going into the directional coupler, which provides another \qty{20}{\dB} of insertion loss before coupling to the JTWPA. An additional isolator is placed after the JTWPA to protect it from signals propagating backward along the readout line. After going through the JTWPA switch, the signal is further amplified by a HEMT amplifier on the \qty{4}{\K} stage, followed by a series of room temperature amplifiers. The amplified signal is then down-converted at the IQ mixer against the LO signal. The digitizer (Keysight M3202A) records both the in-phase (I) and quadrature (Q) components for subsequent analysis.

%-----------------------------------------------
\section{Results}
\label{sec:results}
This section focuses on the characterization results of the aluminum \uMUX prototype. We first compare the measured $S_{21}$ values at different operating points to the multiplexer model introduced in~\cref{sec:model}. This enables us to validate equations~\eqref{eq:squid_dispersion_low_pwr} and \eqref{eq:squid_dispersion_finite_pwr}, improving our understanding of the device properties. We then present noise measurements with a HEMT amplifier, and conclude with methods to further lower the amplifier noise floor.

%%%---------------------------------------------
\subsection{Low RF power characterization} \label{results:low_rf_pwr}

We use low power limit $S_{21}$ measurements to survey the resonator properties, validate the multiplexer model, and extract the circuit parameters needed to predict noise performance. All measurements are done with probe tone power \mbox{$P_{\text{exc}} \approx$~\qty{-104}{dBm}} at device, which produces $\Phi_{\text{rf}} \approx 0.01 \Phi_0 \ll \Phi_0$ at the SQUID.

\begin{figure}[htb]
    \centering
    \includegraphics[width=\columnwidth]{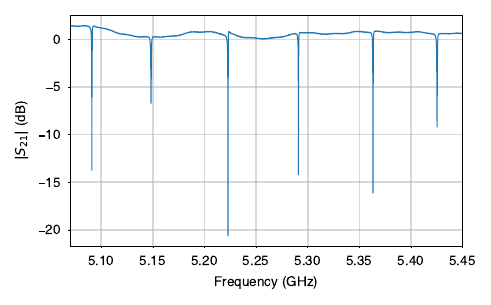}
    \caption{The $|S_{21}|$ trace of all resonances taken at \mbox{$P_{\text{exc}}$ =~\qty{-104}{dBm}}. The $S_{21}$ values in this figure are of the entire RF chain.}
    \label{fig:all_res_trace}
\end{figure}

At a fixed flux bias, we measure the $S_{21}$ of the whole RF chain, as shown in figure~\ref{fig:all_res_trace}. Six resonances are visible. The resonator linewidths range from approximately 350 to \qty{800}{\kHz}, and the frequency spacing between adjacent resonators lies between 50 and \qty{80}{\MHz}.

Using the circle fit algorithm detailed in \cite{probst_efficient_2015}, we extract the quality factors of the resonators. The resulting values of $Q_i$, $Q_c$, and $Q_l$ are summarized in table~\ref{tab:quality_factors}. The internal quality factors $Q_i$ are above $7\times10^4$ for most channels, with the exception of channels 2 and 6. The coupling quality factors $Q_c$ are approximately $1.4\times10^4$ across all channels, placing the device close to the $Q_i \gg Q_c$ regime that is generally desirable. These $Q_i$ values are reduced relative to other resonators constructed from high-quality superconductors, likely due to the geometry near the termination and coupling to many potentially lossy components, such as the SQUID, input coil, and flux bias line. Although it is not entirely clear what limits the resonator $Q_i$, we observe flux-dependent variation of $Q_i$, which can be described by the resistively and capacitively shunted junction (RCSJ) model as found in~\cite{kempf_design_2017}. We discuss this $Q_i$ variation in more detail in Appendix~\ref{appx:Qi}.

\begin{table}[htb]
    \centering
    \begin{tabular}{c c c c c}
    \hline
         Ch & $f_{\text{r}}$ (GHz) & $Q_i$ & $Q_c$ & $Q_l$\\
    \hline
    1 & 5.09 & 68.4k & 13.4k & 11.2k \\
    2 & 5.15 & 16.0k & 15.2k & 7.7k \\
    3 & 5.22 & 138.3k & 13.4k & 12.2k \\
    4 & 5.29 & 75.3k & 17.3k & 14.1k \\
    5 & 5.36 & 80.3k & 12.8k & 11.0k \\
    6 & 5.42 & 27.4k & 12.4k & 8.5k \\
    \hline
    \end{tabular}
    \caption{Quality factors of the resonators at \mbox{$P_{\text{exc}}$ =~\qty{-104}{dBm}}. The quality factors are averaged over a flux quantum of external flux bias.}
    \label{tab:quality_factors}
\end{table}

\begin{figure}[htb]
    \centering
    \includegraphics[width=\columnwidth]{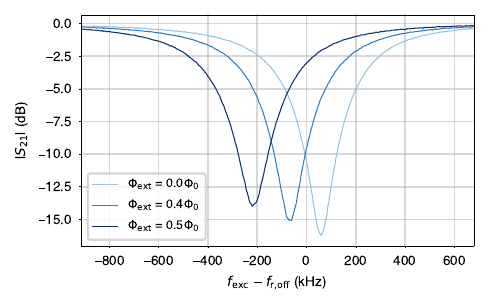}
    \caption{The $|S_{21}|$ traces of channel 1 as a function of flux bias through the SQUID. The $S_{21}$ is rotated with circle fit parameters to the canonical value given by equation \eqref{eq:S_21}. The $x$-axis is shifted by $f_{\text{r,off}}$ (defined in equation~\eqref{eq:squid_dispersion_low_pwr_abbr}) for clarity.}
    \label{fig:single_res_trace}
\end{figure}

\begin{figure}[htb]
    \centering
    \includegraphics[width=\columnwidth]{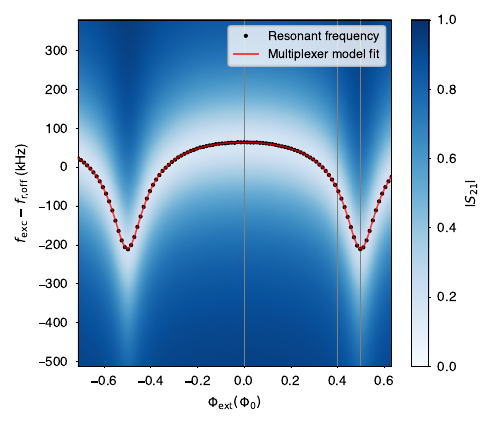}
    \caption{The heatmap shows the resonator $|S_{21}|$ of channel 1 as a function of flux bias through the SQUID ($\Phi_{\text{ext}}$) measured with a VNA. The $S_{21}$ is rotated with circle fit parameters to the canonical value given by equation \eqref{eq:S_21}. The black dots are the resonant frequencies extracted with the circle fit. The red line is the best fit of the multiplexer model in equation \eqref{eq:squid_dispersion_low_pwr_abbr} to the black dots. The gray lines are the flux bias points at which the resonance line shapes are plotted in figure~\ref{fig:single_res_trace}.}
    \label{fig:TES3W12S9_heatmap}
\end{figure}

As detailed in \cref{sec:model}, the resonant frequency shifts periodically with the flux through the SQUID. Figure \ref{fig:single_res_trace} shows this resonance shift between different flux bias points. In order to fit the multiplexer model to experimental data with minimal free parameters, we parametrize equation \eqref{eq:squid_dispersion_low_pwr} as
\begin{equation} \label{eq:squid_dispersion_low_pwr_abbr}
f_{\text{r}}(I_{\text{bias}}) = f_{\text{r,off}} + \Delta f_{\text{r,mod}} \frac{\beta_L \cos{\varphi_{\text{tot}}}}{1+\beta_L \cos{\varphi_{\text{tot}}}}.
\end{equation}
The $\varphi_{\text{tot}}$ in this expression is solved from equation~\eqref{eq:tot_flux} with $\Phi_{\text{ext}}=(I_{\text{bias}}M_{\text{mod}}+\Phi_{\text{off}})$, where $\Phi_{\mathrm{off}}$ is the flux offset within the SQUID loop at zero bias current. Here $I_{\text{bias}}$ is the bias current through flux bias line. The free parameters in this fit are $f_{\text{r,off}}, \Delta f_{\text{r,mod}}, \beta_L, M_{\text{mod}}$, and the flux offset $\Phi_{\text{off}}$. Intuitively, $f_{\text{r,off}}$ represents the total frequency offset of the resonance pattern from zero, and $\Delta f_{\text{r,mod}}$ is related to the peak-to-peak shift of the resonant frequency. For each channel, we measure the resonator $S_{21}$ over a range of $I_{\text{bias}}$ and $f_{\text{exc}}$, which is used to extract $f_r$ at each $I_{\text{bias}}$ through circle fits. Figure \ref{fig:TES3W12S9_heatmap} shows the normalized resonator $|S_{21}|$ of channel 1 at a range of flux bias points with the heatmap and the circle-fitted $f_r$ in black dots. The red line shows the best fit of the multiplexer model to the $f_r$ data points. The agreement between the fit and data confirms the validity of the multiplexer model. The fitted values of $f_{\text{r,off}}, \Delta f_{\text{r,mod}}, \beta_L$, and $M_{\text{mod}}$ for all six channels are shown in table~\ref{tab:mates_fit_params}. These values will be used to extract circuit parameters of the \uMUX.

\begin{table}[htb]
    \centering
    \begin{tabular}{c c c c c}
    \hline
         \small{Ch} & \small{$f_{\text{r,off}}$ (GHz)} & \small{$\Delta f_{\text{r,mod}}$ (kHz)} & \small{$\beta_L$} & \small{$M_{\text{mod}}$ (pH)}\\
    \hline
    1 & 5.091 & 183 & 0.54 & 1.37 \\
    2 & 5.148 & 260 & 0.55 & 1.29 \\
    3 & 5.223 & 152 & 0.54 & 1.25 \\
    4 & 5.291 & 131 & 0.54 & 1.29 \\
    5 & 5.363 & 120 & 0.54 & 1.26 \\
    6 & 5.425 & 177 & 0.54 & 1.40 \\
    \hline
    \end{tabular}
    \caption{Fitted parameters from the multiplexer model at low power limit using equation~\eqref{eq:squid_dispersion_low_pwr}.}
    \label{tab:mates_fit_params}
\end{table}

Under the parametrization of equation~\eqref{eq:squid_dispersion_low_pwr_abbr}, $L_T$ can be calculated from the fitted offset frequency $f_{\mathrm{r,off}}$ with
\begin{equation} \label{eq:L_T_from_fit}
    L_T = \frac{f_0-f_{\text{r,off}}}{4f_0^2}Z_0 - C_c Z_0^2.
\end{equation}
To extract $L_T$, we first require an estimate of the coupling capacitance $C_c$ between the resonator and the feedline. The coupling capacitance can be inferred from the circle-fitted coupling quality factor $Q_c$ using the expression in~\cite{mates_microwave_2011}
\begin{equation}
    Q_c = \frac{\pi}{2 (2 \pi f_0)^2 C_c^2 Z_0^2}.
\end{equation}
Using the measured values of $Q_c$ and assuming \mbox{$Z_0 =$ \qty{55}{\ohm}}, we calculate $C_c$ for each channel. Substituting the extracted $C_c$ together with the previously fitted $f_{\mathrm{r,off}}$ into equation~\eqref{eq:L_T_from_fit} yields $L_T$. The resulting values of $C_c$ and $L_T$ are summarized in table~\ref{tab:circuit_params}.

Similarly, $M_T$ can be calculated from the measured modulation amplitude $\Delta f_{\mathrm{r, mod}}$ with
\begin{equation} \label{eq:M_T_from_fit}
    M_T = \frac{\sqrt{Z_0 L_S \Delta f_{\text{r,mod}}}}{2f_0}.
\end{equation}
To do this, we require an estimate of the SQUID loop inductance $L_S$, which in turn requires an estimate of the Josephson inductance $L_J$ from the junction critical current density $J_c$. We first measure the normal-state resistance $R_n$ of the Josephson junction in the SQUID. Then we calculate $J_c$ using the Ambegaokar--Baratoff relation~\cite{ambegaokar_tunneling_1963}, $J_c = \pi \Delta/2 e R_n A_{\mathrm{JJ}}$, where $\Delta$ is the effective superconducting gap of aluminum, and $e$ is the electron charge. From $J_c$ and the known junction area $A_{\mathrm{JJ}}$, we compute $L_J$ with $L_J= \Phi_0/2\pi J_c A_\text{JJ}$. For our device, we obtain $L_J = $ \qty{387}{pH} with $A_{\mathrm{JJ}} = 1.45 \times$ \qty{0.2}{\um^2} and $J_c =$ \qty{2.93}{\uA/\um^2} based on the process control junctions on the same wafer. Combining this value with the fitted parameter $\beta_L$, we calculate the SQUID loop inductance via $L_S = \beta_L L_J$. Finally, substituting the extracted $L_S$ into equation~\eqref{eq:M_T_from_fit} yields $M_T$. The resulting values of $L_S$ and $M_T$ are listed in table~\ref{tab:circuit_params}, and show reasonable agreement with simulation.

\begin{table}[htb]
    \centering
    \begin{tabular}{c c c c c}
    \hline
         Ch & \small{$C_c$ (fF)} & \small{$L_T$ (pH)}& \small{$L_S$ (pH)} & \small{$M_T$ (pH)} \\
    \hline
    Sim & 4.81 & 86 & 191 & 6 \\
    \hline
    1 & 5.93(1) & 72.7 & 207(5) & 4.34(5) \\
    2 & 5.51(5) & 76.7(2) & 214(5) & 5.18(6) \\
    3 & 5.78 & 71.0 & 210(5) & 3.87(5) \\
    4 & 5.03(1) & 72.5 & 209(5) & 3.53(4) \\
    5 & 5.74(1) & 79.9 & 208(5) & 3.32(4) \\
    6 & 5.78(1) & 70.0 & 210(5) & 4.01(5) \\
    \hline
    \end{tabular}
    \caption{Simulated and measured circuit parameters. The numbers in the parentheses indicate the estimated uncertainty of these values, when significant. The uncertainties are propagated from the circle fit uncertainties and uncertainties of the junction critical current.}
    \label{tab:circuit_params}
\end{table}

\subsection{On-chip power calibration}\label{results:power_calib}

\begin{figure}[htb]
    \centering
    \includegraphics[width=\columnwidth]{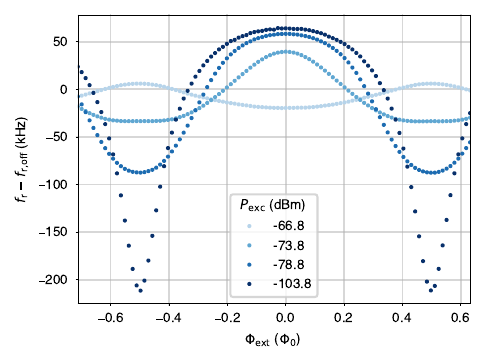}
    \caption{The power dependence of the resonant frequency as a function of flux for channel 1. The range of the resonance shift shrinks with increasing microwave power and eventually flips over at very high powers. Different colors represent data taken at different $P_{\text{exc}}$ as shown in the legends.}
    \label{fig:TES3W12S9_power_bias_scan}
\end{figure}

In order to accurately model the flux sensitivity and estimate the system noise temperature, we require a reliable calibration of the probe tone power at the device. As discussed in \cref{sec:model}, when the multiplexer is probed at higher RF power, the peak-to-peak resonant frequency shift, $\Delta f_{\text{pp}}$, decreases with increasing $P_{\text{exc}}$. Figure~\ref{fig:TES3W12S9_power_bias_scan} shows this behavior in channel 1 of our \uMUX. The dependence of the resonance shift on probe tone power provides a convenient on-chip method for calibrating the microwave attenuation between the RF source and the device.

To perform this calibration, we extract the resonant frequency over a range of probe tone powers and flux bias points. For each probe tone power, we select resonant frequencies at integer, half-integer, and quarter-integer values of $\Phi_{\text{ext}}/\Phi_0$. The finite-power corrections at these operating points are described by a simplified form of equation~\eqref{eq:squid_dispersion_finite_pwr},
\begin{equation} \label{eq:squid_dispersion_finite_pwr_abbr}
    f_r (P_{\text{exc}}, \Phi_{\text{ext}}) \approx f_{\text{r,off}} + \Delta f_{\text{r,mod}} \frac{2 \beta_L}{\varphi_{\text{rf}}} \sum_{m,n} p_{m,n}.
\end{equation}
The normalized AC flux amplitude $\varphi_{\text{rf}}$ in this expression depends on $P_{\text{exc}}$ (see equation~\eqref{eq:Phi_rf}). $P_{\text{exc}}$ is related to the RF source output power by \mbox{$P_{\text{exc}} = A_{\text{in}} P_{\text{src}}$}, where $A_{\text{in}}$ is the total attenuation between the source and the device, and $P_{\mathrm{src}}$ is the power output from the RF source. Consequently, the measured dependence of $f_r$ on $P_{\text{src}}$ and $\Phi_{\text{ext}}$ can be used to extract the attenuation $A_{\text{in}}$. Substituting equation~\eqref{eq:Phi_rf} into equation~\eqref{eq:squid_dispersion_finite_pwr_abbr}, we obtain the final expression used for fitting:
\begin{equation}
    f_r (P_{\text{src}}, \Phi_{\text{ext}}) = f_{\text{r,off}} + \frac{ \beta_L \Phi_0}{4 Q_l M_T} \sqrt{\frac{Q_c Z_0}{\pi A_{\text{in}} P_{\text{src}}}} \sum_{m,n} p_{m,n}.
\end{equation}
In this fit, we fix $f_{\text{r,off}}$, $\Delta f_{\text{r,mod}}$, $\beta_L$, and $M_T$ to the values extracted from the low-power characterization in \cref{results:low_rf_pwr}. We also fix $Q_l$ and $Q_c$ to the values averaged over one flux quantum in the low-power limit (see table~\ref{tab:quality_factors}). With the above parameters constrained by independent measurements, the attenuation $A_{\text{in}}$ is the only free parameter in the fit.

\begin{figure}[htb]
    \centering
    \includegraphics[width=\columnwidth]{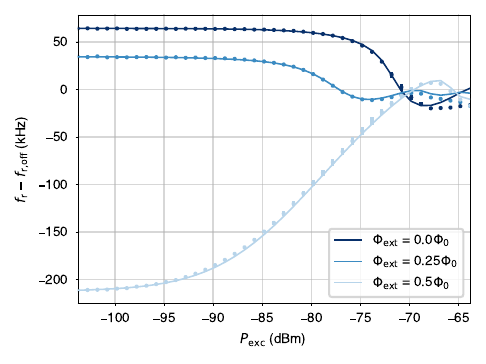}
    \caption{Resonant frequency at different external flux and RF powers of channel 1. The dots are resonant frequency extracted from circle fits to data. The vertical lines on the dots are the error bars, which are smaller than the dots most of the times. The solid lines are best fit of the model to data.}
    \label{fig:TES3W12S9_power_calibration}
\end{figure}

The model reproduces the measured power dependence of the resonance frequency. Figure~\ref{fig:TES3W12S9_power_calibration} compares data and the fit, using channel 1 as an example. The extracted attenuation values for all six channels are consistent within \qty{1}{\dB} (see table~\ref{tab:pwr_attenuation}). Averaging in power units yields a mean attenuation of \qty{-68.82}{\dB}. Prior to installing the device, we independently measured the attenuation of the bare RF drive lines at base temperature using a VNA and obtained \qty{-66.15}{\dB}. The close agreement between these values, given additional losses from switches, connectors, and cryogenic components, validates both the \uMUX model and the circuit parameters extracted in \cref{results:low_rf_pwr}. This calibrated on-chip power is subsequently used in the flux sensitivity and noise temperature analysis in the remainder of this section.

\begin{table}[htb]
    \centering
    \begin{tabular}{c c c c c}
    \hline
         \small{Ch} & \small{$A_{\text{in}}$ (dB)}\\
    \hline
    1 & $-68.60$\\
    2 & $-68.33$\\
    3 & $-68.89$\\
    4 & $-69.17$\\
    5 & $-68.64$\\
    6 & $-69.38$\\
    \hline
    \end{tabular}
    \caption{Best fit values of the attenuation $A_{\text{in}}$ from the \uMUX on-chip power calibration.}
    \label{tab:pwr_attenuation}
\end{table}

%%%---------------------------------------------
\subsection{Open-loop and FRM sensitivity} \label{results:HEMT_noise}

In this section, we quantify the flux sensitivity of the multiplexer in both open-loop and flux-ramping operations and compare the measured noise performance to model predictions. We measure the NSD at the optimal readout power for FRM, with a HEMT amplifier as the first-stage amplifier. The optimal power is determined by the trade-off between the linear increase of the signal-to-noise ratio (SNR) with $P_{\mathrm{exc}}$ in the low-power limit and the reduction of SQUID responsivity at higher powers. According to \cite{mates_simultaneous_2017} and \cite{schuster_simulation_2023}, the optimal probe tone power for FRM corresponds to an RF-induced flux amplitude of \mbox{$\Phi_{\mathrm{rf}} \approx 0.3\,\Phi_0$}. We use the power calibration from \cref{results:power_calib} to calculate the optimal power for each channel and use them for all measurements within this section.

\begin{figure}[htb]
    \centering
    \includegraphics[width=\columnwidth]{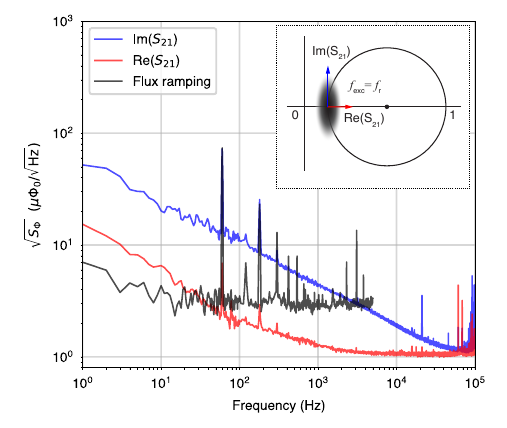}
    \caption{Open-loop and FRM noise spectra of channel 1 in equivalent flux units. All three noise traces were taken with optimal probe tone power $P_{\text{exc}}\approx$~\qty{-77}{dBm}. The inset shows the $S_{21}$ circle in the complex plane for a canonical notch type resonator. The ellipse represents the open-loop noise blob when probed on resonance. The blue and red arrows show the directions along which we project the $S_{21}$ data. The noise spectral density traces with corresponding colors are obtained in $\text{Im}(S_{21})$ and $\text{Re}(S_{21})$ quadrature respectively. The black line shows the noise spectral density when operating in FRM mode. The \qty{60}{\Hz} noise and its harmonics are intrinsic to the function generator of the flux-ramp signal.}
    \label{fig:HEMT_only_PSD}
\end{figure}

The open-loop NSDs of channel 1 measured at the optimal probe tone power are shown in figure~\ref{fig:HEMT_only_PSD}. The resonator was probed on resonance, and the noise was measured at the sensitive flux bias point where the transfer function $\partial \mathrm{Im}(S_{21})/\partial \Phi_{\mathrm{ext}}$ is maximized. The inset shows an illustration of imaginary and real quadratures into which the noise is projected. The imaginary quadrature, $\mathrm{Im}(S_{21})$, is sensitive to resonance frequency fluctuations and therefore to flux changes. The real quadrature, $\mathrm{Re}(S_{21})$, is largely insensitive to shifts in resonant frequency and primarily reflects dissipative noise and amplifier noise. At frequencies below approximately \qty{100}{\kHz}, the $\mathrm{Im}(S_{21})$ NSD exhibits significant $1/f$ scaling, whereas the $\mathrm{Re}(S_{21})$ NSD remains flat above \qty{10}{\kHz}. Since the frequency-sensitive quadrature shows more pronounced $1/f$ noise, we attribute it primarily to two-level system (TLS) fluctuations in the resonator, which manifest as resonance frequency noise~\cite{gao_noise_2007, kumar_temperature_2008}. The fact that this $1/f$ noise is greatly reduced in the FRM NSD also supports that it mainly originates downstream from the SQUID. Flux noise within the SQUID loop can add a subdominant contribution to the low frequency noise since it also has a $1/f$ spectral feature \cite{wellstood_low-frequency_1987, rower_evolution_2023}. At higher frequencies, the noise levels in the real and imaginary quadratures converge, indicating that both are dominated by white noise from the first-stage amplifier. We model the full $\text{Im}(S_{21})$ noise spectrum with
\begin{equation}
S^{\mathrm{Im}}_{\Phi}(f) = S^{\mathrm{Im}}_{\Phi,w} + f^{-a} \, S^{\mathrm{Im}}_{\Phi,1/f}(1\,\mathrm{Hz}),
\end{equation}
where $S^{\mathrm{Im}}_{\Phi,w}$ denotes the white amplifier noise level and $S^{\mathrm{Im}}_{\Phi,1/f}(1\,\mathrm{Hz})$ characterizes the $1/f$ noise amplitude at \qty{1}{Hz}. When fitting this expression to the NSD, we fix $S^{\mathrm{Im}}_{\Phi,w}$ to the $\mathrm{Re}(S_{21})$ quadrature noise averaged over 10--\qty{40}{\kHz}, as this provides a direct measurement of the amplifier-limited noise. The best fit values for all channels are summarized in table~\ref{tab:open_loop_noise}. To remain consistent with other literature, we use $\sqrt{S^{\mathrm{Im}}_{\Phi,w}}$ as the flux sensitivity of our device. The measured flux sensitivities range between 1--1.5~\uPhirtHz across the 6 channels. This is comparable with recently published niobium devices, which have open-loop flux sensitivities of 0.45~\uPhirtHz~\cite{malnou_improved_2023} and 0.7~\uPhirtHz~\cite{neidig_full-scale_2025}.

Using the circuit parameters extracted in sections~\ref{results:low_rf_pwr} and \ref{results:power_calib} and assuming a system noise temperature $T_{\text{sys}}$ (referenced to the \uMUX output) equal to the specified HEMT noise temperature, $T_{\text{HEMT}} = 2.8~\mathrm{K}$, we calculate the expected flux sensitivity from the multiplexer model (see equation~\eqref{eq:OL_noise_model}). The predicted values are listed in table~\ref{tab:open_loop_noise}. Overall, they show reasonable agreement with measured noise levels, supporting the validity of the extracted circuit parameters and the noise model. There is a systematic trend that the predicted noise level is slightly lower than the measured values. This discrepancy can be attributed to the loss between the \uMUX output and the HEMT input. In particular, a transmission factor of $A_{\text{out}} \le 1$ increases the effective system noise temperature by $T_{\text{sys}} = T_{\text{HEMT}}/A_{\text{out}}$. Since the measured flux sensitivity $\sqrt{S_{\Phi,w}}$ scales with $T_{\text{sys}}$, using $T_{\text{HEMT}}$ in the model leads to a slight underestimation of the noise floor compared to the measurements.

\begin{table}[htb]
    \centering
    \begin{tabular}{c c ccc}
    \hline
         \multirow{2}{*}{Ch} & predicted & \multicolumn{3}{c}{measured} \\
         \cline{2-5}
         & $\sqrt{S_{\Phi,w}}$ & $\sqrt{S_{\Phi,w}}$ & \small{$\sqrt{S_{\Phi_, 1/f} (1 \text{Hz})}$} & $a$ \\
    \hline
    1 & 0.89 & $1.06 (4)$ & 58.84 & 0.77\\
    2 & 1.31 & $1.46 (9)$ & 66.34 & 0.85\\
    3 & 0.84 & $0.96 (6)$ & 76.05 & 0.81\\
    4 & 0.87 & $1.10 (5)$ & 41.91 & 0.80\\
    5 & 1.10 & $1.17 (12)$ & 93.11 & 0.77\\
    6 & 1.27 & $1.26 (19)$ & 124.80 & 0.84\\
    \hline
    \end{tabular}
    \caption{Predicted and measured open-loop flux equivalent noise in units of \uPhirtHz. For the measured $\sqrt{S_{\Phi,w}}$ values, the numbers in the parentheses indicate the estimated uncertainty. See Appendix~\ref{appx:uncertainties} for more details on the uncertainty estimation. Channels 5 and 6 have significantly higher uncertainty than the other channels. This is because these two channels have larger low frequency noise, which increases the uncertainty of the measured transfer function $d\Phi/d\mathrm{Im}(S_{21})$.}
    \label{tab:open_loop_noise}
\end{table}

The FRM noise is measured with a ramp frequency of $f_{\mathrm{ramp}} = $~\qty{10}{\kHz} and a peak-to-peak ramp amplitude of approximately $3.5\,\Phi_0$. The modulation frequency $f_{\text{mod}}$ is thus around \qty{35}{kHz}. To avoid contamination from transients, the first $20\%$ of the data within each ramp interval is discarded during demodulation, resulting in a duty cycle of \mbox{$\alpha = 0.8$}. In \cref{model:sensitivity}, we discussed how resonator TLS noise can be largely suppressed by FRM. It is clear in figure~\ref{fig:HEMT_only_PSD} that the FRM noise spectrum at frequencies below \qty{10}{\kHz} is no longer overwhelmed by $1/f$ noise. Thus the flux sensitivity is primarily limited by amplifier noise plus any excess noise at $f_{\text{mod}}$. The measured flux sensitivities with FRM $\sqrt{S_{w, \text{FRM}}}$ range over 3--4~\uPhirtHz~for our \uMUX (see table~\ref{tab:FRM_noise}). This is roughly within a factor of two of the recent niobium devices, which have FRM sensitivities of 1.6~\uPhirtHz~\cite{malnou_improved_2023} and 1.4~\uPhirtHz~\cite{neidig_full-scale_2025}. With a measured $M_{\mathrm{in}} = 251~\mathrm{pH}$, the flux sensitivity of our device corresponds to an input current sensitivity of 24--33~$\mathrm{pA}/\sqrt{\mathrm{Hz}}$.

\begin{table}[htb]
    \centering
    \begin{tabular}{c c}
    \hline
         \small{Ch} & \small{$\sqrt{S_{w, \text{FRM}}}$ (\uPhirtHz)}\\
    \hline
    1 & $3.1 (2)$\\
    2 & $3.9 (2)$\\
    3 & $2.9 (1)$\\
    4 & $2.9 (2)$\\
    5 & $3.6 (2)$\\
    6 & $3.9 (2)$\\
    \hline
    \end{tabular}
    \caption{FRM flux sensitivity. The numbers in the parentheses indicate the estimated uncertainty. See Appendix~\ref{appx:uncertainties} for more details on uncertainty estimation.}
    \label{tab:FRM_noise}
\end{table}

Lastly, we checked the linearity of the flux reconstruction in FRM mode. The deviation from a perfectly linear response is around 1 $\mathrm{m}\Phi_0$ (see Appendix~\ref{appx:linearity} for details).

%%%---------------------------------------------
\subsection{JTWPA sensitivity}
\label{results:TWPA}

For CE$\nu$NS and light dark matter experiments, achieving low energy thresholds requires detectors with correspondingly low noise. The intrinsic detector noise of a Q-Array style detector (solid state target with phonon-sensing TESs) can be as low as \mbox{10--15 $\mathrm{pA}/\sqrt{\mathrm{Hz}}$}~\cite{augier_results_2023, chen_design_2025}. To prevent the readout chain from limiting the overall performance, the \uMUX noise must be reduced to a comparable or lower level. Several approaches can be pursued to reach this goal, including optimizing the multiplexer circuit parameters, increasing the coupling of the input coil to the SQUID $M_{\text{in}}$, and using first-stage amplifiers with lower noise. In this section, we demonstrate a reduction in readout noise by incorporating a JTWPA fabricated at Lincoln Laboratory~\cite{macklin_nearquantum-limited_2015} as the first-stage amplifier in the readout chain.

\begin{figure}[htb]
    \centering
    \includegraphics[width=\columnwidth]{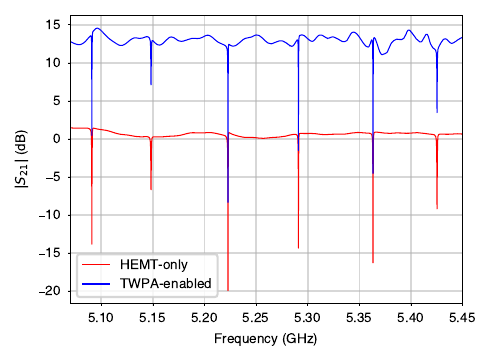}
    \caption{Transmission coefficient of the RF chain with and without JTWPA. Measured at \mbox{$P_{\text{exc}} =$~\qty{-84}{dBm}}.}
    \label{fig:TWPA_gain}
\end{figure}

The JTWPA is installed in between a pair of switches right after the device on the mixing chamber (see figure~\ref{fig:setup_schematic}). The JTWPA pump tone has a frequency of \qty{7.9}{\GHz} and a power around \qty{-58}{dBm} at device. Figure~\ref{fig:TWPA_gain} shows the transmission coefficient of the RF chain with the JTWPA and with the JTWPA bypassed. This demonstrates that our JTWPA has a gain of around \qty{13}{\dB} in the frequency region of the \uMUX (5--\qty{5.5}{\GHz}).

\begin{figure}[htb]
    \centering
    \includegraphics[width=\columnwidth]{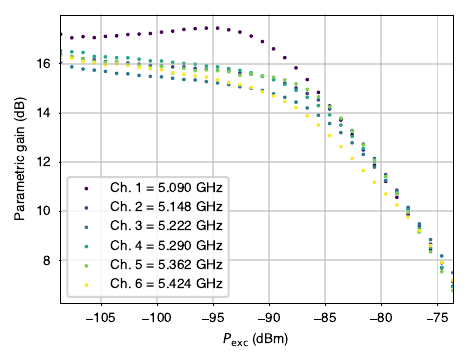}
    \caption{Parametric gain of the JTWPA near the \uMUX resonances at different probe tone powers. The parametric gain is defined as the ratio of the transmissions in JTWPA pump on and off configurations.}
    \label{fig:TWPA_compression}
\end{figure}

Our JTWPA is not optimized to operate at the optimal probe tone power of the present multiplexer device (\mbox{$\approx$~\qty{-76}{dBm}}). Therefore, we must select a power that somewhat preserves the \uMUX noise sensitivity while avoiding substantial gain compression in the JTWPA. By scanning $P_{\mathrm{exc}}$, we characterized the gain compression profile of the JTWPA near the resonator frequencies of this device (see figure~\ref{fig:TWPA_compression}). Using the attenuation extracted in section~\ref{results:power_calib} and assuming negligible additional losses in the circuit, we determine a \qty{1}{\dB} compression point ($P_{-1\text{dB}}$) between $-90$ and \qty{-85}{dBm}. As anticipated, this compression threshold lies below the optimal operating power of a single \uMUX channel. As a compromise, we chose to operate at \mbox{$P_{\mathrm{exc}} = -84~\mathrm{dBm} \approx P_{-1\text{dB}}$}. We are also aware that operating near the nonlinear regime of the JTWPA can generate third-order intermodulation products, leading to crosstalk between tones~\cite{mates_crosstalk_2019}. From the measurements described in Appendix~\ref{appx:TWPA_IP3}, we obtain the input-referred third-order intercept point \mbox{$\mathrm{IIP3} \approx$ \qty{-82.5}{dBm}}, which is comparable to our operating power. While this does not affect the present measurements since we probe one channel at a time, it highlights the need for higher dynamic range amplifiers for multi-channel multiplexed operation.

\begin{figure}[htb]
    \centering
    \includegraphics[width=\columnwidth]{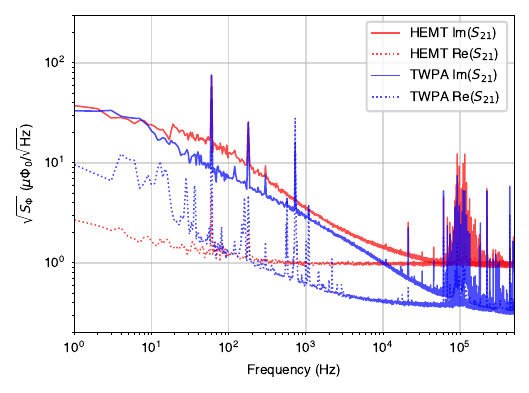}
    \caption{Open-loop noise of channel 1 with and without the JTWPA. The \qty{60}{\Hz} noise and its harmonics are intrinsic to the function generator of the flux-ramp signal.}
    \label{fig:TWPA_open_loop_noise}
\end{figure}

\begin{table*}[htb]
    \centering
    \begin{tabular}{c cc cc cc}
    \hline
    \multirow{2}{*}{Ch} & \multicolumn{2}{c}{$\sqrt{S_{\Phi,w}}$ (\uPhirtHz)} & \multicolumn{2}{c}{$T_{\text{sys}}$ (K)} & \multicolumn{2}{c}{$\sqrt{S^{\text{Im}}_{\text{FRM}}}$ (\uPhirtHz)} \\
    \cline{2-7}
    & HEMT-only & JTWPA-enabled & HEMT-only & JTWPA-enabled & HEMT-only & JTWPA-enabled \\
    \hline
    1 & 0.98 & 0.33 & 3.97 & 0.42 & 3.98 & 2.73 \\
    2 & 1.56 & 0.55 & 4.06 & 0.61 & 6.29 & 3.58 \\
    3 & 0.89 & 0.36 & 3.64 & 0.69 & 3.90 & 2.68 \\
    4 & 1.00 & 0.43 & 3.04 & 0.53 & 4.40 & 2.80 \\
    5 & 1.24 & 0.54 & 3.83 & 0.68 & 6.21 & 5.46 \\
    6 & 1.23 & 0.62 & 2.87 & 0.76 & 5.94 & 3.83 \\
    \hline
    \end{tabular}
    \caption{Flux sensitivities of open-loop and FRM operations measured at \mbox{$P_{\text{exc}}\approx$~\qty{-84}{dBm}}. The system noise temperatures $T_{\text{sys}}$ are calculated from the open-loop noise spectral density for both HEMT-only and JTWPA-enabled configurations.}
    \label{tab:TWPA_noise}
\end{table*}

We measured both the open-loop and flux-ramping noise. The open-loop noise is measured on resonance at the sensitive flux bias point. Figure~\ref{fig:TWPA_open_loop_noise} shows the NSDs of channel~1 with and without the JTWPA. The flux sensitivity improves from 0.98~\uPhirtHz to 0.33~\uPhirtHz after adding the JTWPA. The corresponding values for all channels are summarized in table~\ref{tab:TWPA_noise}, demonstrating a consistent factor of 2--3 improvement in sensitivity across channels. To quantify this improvement in terms of amplifier performance, we convert the measured white noise levels from NSDs to an equivalent system noise temperature. We calculate $T_{\text{sys}}$ from data using \mbox{$S^{\text{Re}} = k_B T_{\text{sys}} / P_{\text{exc}}$}, where $S^{\text{Re}}$ is obtained by averaging the NSD of $\text{Re}(S_{21})$ between 10 and \qty{40}{\kHz}. The values of $P_{\text{exc}}$ are determined using the fitted attenuation values in table~\ref{tab:pwr_attenuation} for each channel. Table~\ref{tab:TWPA_noise} lists the extracted noise temperatures for the JTWPA-enabled and HEMT-only configurations. Averaging over all six channels, we obtain \mbox{$T_\text{sys} = 3.6 \pm 0.6~\mathrm{K}$} without the JTWPA and \mbox{$0.62 \pm 0.17~\mathrm{K}$} with the JTWPA, where the quoted uncertainties are conservatively estimated from the spread of $T_{\text{sys}}$ values across channels. This measured performance is roughly consistent with the values reported for JTWPAs of similar design, such as $0.602~\pm$ \qty{0.015}{\K} in~\cite{macklin_nearquantum-limited_2015} and $1.38~\pm$ \qty{0.08}{\K} in~\cite{bartram_dark_2023}. For comparison, the standard quantum limit (SQL) that corresponds to a noise energy of one photon, gives an equivalent noise temperature $T_{\mathrm{SQL}} = \hbar \omega / k_B$. For our operating frequencies of 5--\qty{5.5}{\GHz}, $T_{\mathrm{SQL}} \approx 0.25~\mathrm{K}$. The readout chain therefore operates at approximately 14 times the SQL without the JTWPA and about 2.5 times the SQL with the JTWPA. This demonstrates that the aluminum \uMUX readout chain can approach quantum-limited sensitivity when paired with a TWPA that has sufficient dynamic range.

\begin{figure}[htb]
    \centering
    \includegraphics[width=\columnwidth]{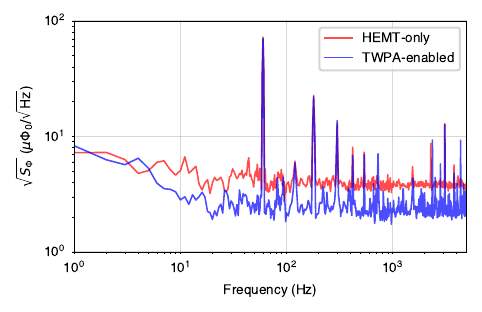}
    \caption{FRM noise of channel 1 with and without the JTWPA. The \qty{60}{\Hz} noise and its harmonics are intrinsic to the function generator of the flux-ramp signal.}
    \label{fig:TWPA_FRM_noise}
\end{figure}

Adding the JTWPA also improves the FRM sensitivity. Figure~\ref{fig:TWPA_FRM_noise} shows the flux-ramping NSD of channel~1 with and without the JTWPA. The white noise level decreases from 3.98~\uPhirtHz to 2.73~\uPhirtHz (see table~\ref{tab:TWPA_noise}). Across most channels, there is an improvement by a factor of 1.5--1.7. The exception is channel~5, where $1/f$ noise remains dominant at the flux-ramp modulation frequency used in this measurement (\mbox{$\approx$~\qty{35}{\kHz}}). Overall, the improvement in FRM sensitivity is less pronounced than that observed in the open-loop white noise. This is because $1/f$ noise dominates over amplifier noise at the modulation frequency, which cannot be suppressed by FRM. We also note that the HEMT-only flux sensitivity reported here is higher than the values listed in table~\ref{tab:FRM_noise}, as the measurements were performed at a probe power approximately \qty{10}{\dB} below the optimal multiplexer operating point in order to accommodate the limited dynamic range of the JTWPA. For realistic operation with larger multiplexing factors, amplifiers with both low noise and higher dynamic range (on the order of \qty{-60}{dBm}) are required, such as the kinetic inductance TWPA (KI-TWPA) ~\cite{malnou_three-wave_2021, faramarzi_48_2024}. Tone-tracking can further help bridge the gap between the high \uMUX readout power and the limited dynamic range of quantum-limited amplifiers~\cite{yu_slac_2023, groh_demonstration_2025}. This is achieved by dynamically adjusting the probe tone frequency to follow the resonance shifts during flux-ramping, thereby reducing the effective power at the amplifier input.

%-----------------------------------------------
\section{Conclusion}
\label{sec:conclusion}

We have designed, fabricated, and characterized a six-channel aluminum \uMUX prototype for TES readout in the \ricochet experiment, demonstrating that an aluminum-based fabrication process can achieve performance comparable to existing niobium devices. By fitting the measured $S_{21}$ response, we extract circuit parameters such as the SQUID loop inductance and resonator termination inductance, which agree reasonably well with electromagnetic simulations. Noise measurements yield open-loop flux sensitivity of 1--1.5~\uPhirtHz. This is consistent with model calculations using measured circuit parameters and the HEMT noise temperature. In FRM operation, low-frequency noise is suppressed, resulting in a flux sensitivity of 3--4~\uPhirtHz, which corresponds to an input current sensitivity at the level of $24$--$33~\mathrm{pA}/\sqrt{\mathrm{Hz}}$. This performance is close to sufficient for our detector readout and comparable to that achieved by existing niobium devices. We further demonstrate noise reduction using a JTWPA as the first-stage amplifier. With the JTWPA, the open-loop flux sensitivity is reduced by a factor of 2 to 3. This corresponds to an effective system noise temperature around \qty{0.6}{\K}, which lies within 2.5 times the SQL. The improvement in FRM sensitivity is more modest, limited by residual $1/f$ noise at the modulation frequency and by the dynamic range of the JTWPA.

There remain several avenues for further improvement. The present single-loop SQUID design is susceptible to fluctuations in the local magnetic field. Implementing a clover-shaped gradiometric geometry would suppress the uniform and first-order magnetic field noise as has been implemented in~\cite{kempf_design_2017, mates_microwave_2011}. Higher per-channel bandwidth is also desirable to enable faster sampling. This can be achieved by increasing the capacitive coupling between the resonator and the RF feedline to reduce the resonator rise time. In addition, further gains in flux sensitivity are possible through optimization of the \uMUX circuit parameters. For example, simulations in \cite{schuster_simulation_2023} suggest that increasing the ratio between the resonance linewidth and the peak-to-peak resonant frequency shift, $\eta$, can improve the FRM flux sensitivity. Since $\eta$ scales with $M_T$, increasing $M_T$ provides a practical route toward improved sensitivity. This also puts an emphasis on developing dedicated simulations of the FRM sensitivity to enable systematic parameter optimization rather than empirical tuning. Finally, after exhausting on-chip design improvements, higher dynamic range TWPAs and tone-tracking will be essential for achieving further reductions of the amplifier noise.

Overall, this work establishes aluminum $\mu$MUX devices as a viable and scalable readout technology for low-noise TES arrays, with clear paths toward further sensitivity improvements.

\appendix
\section{Flux dependence of $Q_i$} \label{appx:Qi}

\begin{figure}[htb]
    \centering
    \includegraphics[width=\columnwidth]{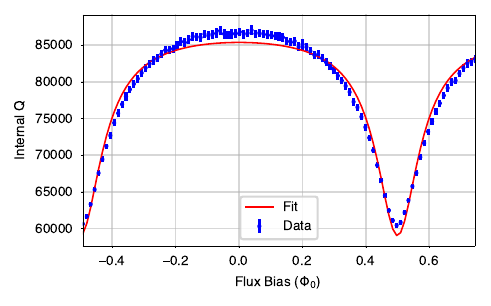}
    \caption{Variation of $Q_i$ with external flux bias on channel 5 measured at \mbox{$P_{\text{exc}}\approx$~\qty{-104}{dBm}}. The blue points are $Q_i$ values obtained from circle fits. The red line is best fit to equation \eqref{eq:Qi_tot}.}
    \label{fig:Qi_fit}
\end{figure}

We observe a periodic variation of the internal quality factor $Q_i$ with applied flux (see figure~\ref{fig:Qi_fit}). The maximum and minimum values of $Q_i$ of all channels are listed in table~\ref{tab:R_sg_fits}. In an ideal dissipationless SQUID, $Q_i$ would be independent of flux. Therefore the flux dependence of $Q_i$ indicates a potential dissipation mechanism associated with the RF SQUID. To identify the dominant source of this dissipation, we consider two possible mechanisms: loss through the flux bias line and dissipation originating from the Josephson junction itself.

\begin{table}[htb]
    \centering
    \begin{tabular}{c c c c c}
    \hline
         Ch & $Q_{\text{i,min}}$ & $Q_{\text{i,max}}$ & $Q_{\text{i,off}}$ & $R_J$ ($\Omega$) \\
    \hline
    1 & 53.3k & 74.2k & 74.8k & 364 \\
    2 & 9.6k & 18.6k & 19.5k & 56 \\
    3 & 104.3k & 148.7k & 154.6k & 536 \\
    4 & 56.6k & 82.6k & 84.5k & 233 \\
    5 & 60.4k & 87k & 89.3k & 236 \\
    6 & 19.2k & 30.3k & 31.7k & 93 \\
    \hline
    \end{tabular}
    \caption{The maximum and minimum $Q_i$ values measured at \mbox{$P_{\text{exc}}\approx$~\qty{-104}{dBm}}. We also show the best fit values of $Q_{\text{i,off}}$ and $R_J$ from fitting equation~\eqref{eq:Qi_tot} to data.}
    \label{tab:R_sg_fits}
\end{table}

The first possible dissipation pathway is through the flux bias line. The wiring to room temperature contributes approximately $R_F = 23~\Omega$ of resistance. The LC low-pass $\pi$-filter on this line introduces an inductance of $L_F = 3300~\mathrm{nH}$. Following the analysis in~\cite{mates_microwave_2011}, the upper limit on $Q_i$ imposed by this dissipation channel is
\begin{gather}
Q_i =
\frac{\pi Z_0}{4 R_F}
\frac{\left(2L_S(L_{\mathrm{mod}} + L_F) + M_{\mathrm{mod}}^2\right)^2}
{M_T^2 M_{\mathrm{mod}}^2 + 4L_S\left(L_S M_p^2 + M_{\mathrm{mod}} M_T M_p\right)},
\label{eq:Qi_fluxline}
\end{gather}
where $Z_0$ is the characteristic impedance of the quarter-wave resonator, which is $55~\Omega$, $M_p$ is the mutual inductance between the resonator termination and flux bias line, and $L_{\text{mod}}$ is the self inductance of the flux bias line. From electromagnetic simulations in table~\ref{tab:COMSOL_sim} we obtain $L_{\mathrm{mod}} = 152~\mathrm{pH}$ and $M_p = 1~\mathrm{pH}$. By averaging the measurement results across all channels in~\cref{results:low_rf_pwr}, we obtained $L_S = 209.7~\mathrm{pH}$, $M_{\mathrm{mod}} = 1.31~\mathrm{pH}$, and $M_T = 3.85~\mathrm{pH}$. Substituting all these values into equation~\eqref{eq:Qi_fluxline}, we find $Q_i \approx 2.0 \times 10^{13}$. This value exceeds the measured $Q_i$ (maximum of order $10^5$) by eight orders of magnitude, demonstrating that dissipation through the flux bias line is negligible in our device.

The second possible source of dissipation originates from the Josephson junction in the RF SQUID. We can model the junction with the RCSJ model, in which the ideal Josephson inductance is shunted by a capacitance $C_{\mathrm{JJ}}$ and a resistance $R_J$. This resistance $R_J$ quantifies the amount of resistive loss associated with the SQUID. The junction impedance is therefore
\begin{gather}
Z_{\mathrm{JJ}} =
\left(
i \omega C_{\mathrm{JJ}}
+ \frac{1}{i \omega L_J \sec \phi}
+ \frac{1}{R_J}
\right)^{-1},
\label{eq:ZJJ}
\end{gather}
where $\phi$ is the superconducting phase difference across the junction. It can be replaced with $\varphi_{\mathrm{tot}}$ in the following calculations. The effective impedance of the resonator termination becomes
\begin{gather}
Z_{\mathrm{eff}} = i \omega
\left(
L_T - \frac{M_T^2}{L_S + Z_{\mathrm{JJ}}/i\omega}
\right).
\label{eq:Z_eff}
\end{gather}

For our device, $C_{\mathrm{JJ}} \approx 15~\mathrm{fF}$ and the resonator frequency is $\omega/2\pi \approx$~\qty{5}{\GHz}, placing us in the regime $\omega C_{\mathrm{JJ}} \ll 1~\Omega^{-1}$. We also assume $R_J \gg$~\qty{1}{\ohm}. In this limit, we can simplify equation~\eqref{eq:Z_eff} and derive the quality factor from its real part
\begin{gather}
Q_{\mathrm{JJ}} =
\frac{\pi Z_0}{4 \operatorname{Re}(Z_{\mathrm{eff}})}
=
\frac{\pi Z_0}{4}
\frac{(1 + \beta_L \cos \varphi_{\mathrm{tot}})^2 R_J}
{(\omega M_T)^2}.
\label{eq:QJJ}
\end{gather}
The detailed derivation can be found in~\cite{mates_microwave_2011}. The flux dependence enters through the term $(1 + \beta_L \cos \varphi_{\mathrm{tot}})^2$, producing a periodic modulation of the dissipative contribution to $Q_i$ with flux. Finally, we model the total internal quality factor as the parallel combination of a flux-independent contribution $Q_{\mathrm{i,off}}$ and the junction-induced contribution $Q_{\mathrm{JJ}}$ ~\cite{kempf_design_2017}
\begin{gather}
Q_i =
\left(
\frac{1}{Q_{\mathrm{i,off}}}
+
\frac{1}{Q_{\mathrm{JJ}}}
\right)^{-1}.
\label{eq:Qi_tot}
\end{gather}

We fit equation~(\ref{eq:Qi_tot}) to the measured $Q_i(\Phi_{\mathrm{ext}})$ data for each channel, treating $Q_{\mathrm{i,off}}$ and $R_J$ as free parameters while fixing $Z_0$, $\omega$, $M_T$, and $\beta_L$ to the values extracted in~\cref{results:low_rf_pwr}. An example fit is shown in figure~\ref{fig:Qi_fit}. The extracted $R_J$ values range from \qty{50}{\ohm}--\qty{500}{\ohm} (see table~\ref{tab:R_sg_fits}), similar to other experiments that also modeled this loss with the RCSJ model~\cite{kempf_design_2017}. This demonstrates that the RCSJ model for dissipation accurately captures the variation of $Q_i$ observed. However, uncovering the actual mechanism of this resistive loss still requires further investigation.

\section{Uncertainty estimation} \label{appx:uncertainties}
In this appendix, we describe the procedure used to estimate the uncertainties of the measured open-loop and FRM flux sensitivities.

The open-loop flux sensitivity, $\sqrt{S_{\Phi,w}}$, is obtained by averaging the $\mathrm{Re}(S_{21})$ noise spectrum over the frequency range 10--\qty{40}{\kHz}. The statistical uncertainty of this averaged value can be approximated as \mbox{$\sigma_{\mathrm{avg}} = \sigma_{\mathrm{raw}}/\sqrt{N}$}, where $N$ is the total number of independent frequency bins included in the average. In our analysis, we average over 50 raw NSD traces and approximately 3000 frequency bins, resulting in a statistical uncertainty below 1\%. This contribution is therefore negligible compared to other sources of uncertainty.

We next consider systematic uncertainties. The flux sensitivity is calculated from real-quadrature noise as
\begin{equation}
    \sqrt{S_{\Phi,w}} = \sqrt{S^{\text{Re}}} \frac{d\Phi}{d\,\mathrm{Im}(S_{21})}.
\end{equation}
The transfer function $H \equiv d\Phi/d\,\mathrm{Im}(S_{21})$ at the operating point is extracted from a calibration dataset, in which $S_{21}$ is measured over a range of $f_{\mathrm{exc}}$ and $\Phi_{\mathrm{ext}}$ values. Because this calibration is performed with finite averaging time, the extracted values of $H$ exhibit statistical fluctuations. We estimate the uncertainty in $H$ by evaluating its spread among neighboring $f_{\mathrm{exc}}$ and $\Phi_{\mathrm{ext}}$ points near the operating bias. Assuming that the uncertainty in $H$ dominates the overall uncertainty budget, the relative uncertainty of $\sqrt{S_{\Phi,w}}$ is then equal to that of $H$. The resulting relative uncertainties for each channel are summarized in table~\ref{tab:uncertainties}. Since these systematic uncertainties far exceed the statistical contribution, we use them as the total uncertainty on $\sqrt{S_{\Phi,w}}$.

\begin{table}[htb]
    \centering
    \begin{tabular}{c c c c c}
    \hline
         Ch & $\sigma_{\text{OL}}$ & $\sigma_{\text{FRM, stats}}$ & $\sigma_{\text{FRM, sys}}$ & $\sigma_{\text{FRM, tot}}$\\
    \hline
    1 & 4.2 & 1.3 & 3.3 & 3.5\\
    2 & 6.0 & 0.8 & 2.8 & 2.9\\
    3 & 5.8 & 1.4 & 4.3 & 4.5\\
    4 & 4.2 & 1.4 & 2.3 & 2.7\\
    5 & 10.0 & 0.9 & 1.4 & 1.7\\
    6 & 15.2 & 0.8 & 1.3 & 1.5\\
    \hline
    \end{tabular}
    \caption{Estimated one-sided relative uncertainties of open-loop and FRM noise floors at optimal operating power. All units in \%.}
    \label{tab:uncertainties}
\end{table}

The FRM sensitivity is determined in an analogous manner. The measured FRM NSD is fitted to
\begin{gather}
    S_{\mathrm{FRM}}(f) = S_{\mathrm{FRM},w} + f^{-n} \, S_{\mathrm{FRM},1/f}(1~\mathrm{Hz}),
\end{gather}
and the fitted parameter $S_{\mathrm{FRM},w}$ is taken as the white noise level. The statistical uncertainty on $S_{\mathrm{FRM},w}$ is estimated from the covariance matrix of the fit parameters. Here each NSD is obtained by averaging 10 raw traces, each containing 680 frequency bins. Taking into account the averaging, the resulting relative statistical uncertainty is approximately 1\% (see table~\ref{tab:uncertainties}).

\begin{figure}[t]
    \centering
    \includegraphics[width=\columnwidth]{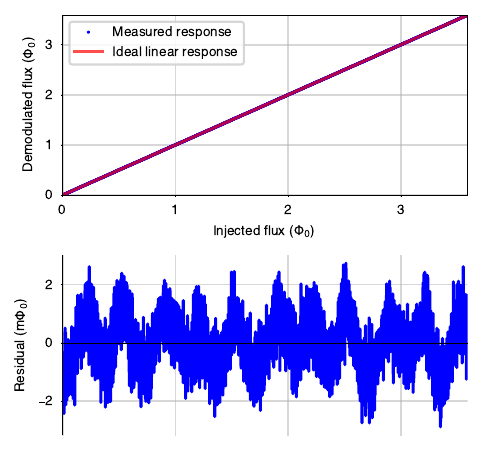}
    \caption{Linearity check of channel 1 at its optimal probe tone power. The top plot shows the demodulated flux as a function of the injected flux. The bottom plot shows the residual of the measured response against a perfectly linear response.}
    \label{fig:HEMT_only_linearity}
\end{figure}

To estimate the systematic uncertainty, we inject a flux signal whose amplitude is independently calibrated using DC measurements. Specifically, we apply a \qty{1}{\Hz} triangle wave through the flux bias line and reconstruct its amplitude using FRM demodulation. The difference between the reconstructed amplitude $A_{\mathrm{FRM}}$ and the DC-calibrated amplitude $A_{\mathrm{DC}}$ is used to quantify the systematic uncertainty. The relative systematic uncertainty on $S_{\mathrm{FRM},w}$ is taken to be $|A_{\mathrm{FRM}} - A_{\mathrm{DC}}|/{|A_{\mathrm{DC}}|}$,
with values summarized in table~\ref{tab:uncertainties}. In principle, this systematic contribution can be calibrated out if the signal pulse shape is well known. Since this calibration was not performed in this work, we retain this contribution to the uncertainty in this work. Then the statistical and systematic uncertainties are added in quadrature to obtain the total uncertainty. Since the calibration is performed with a \qty{1}{\Hz} triangle wave, which may not represent the response to actual detector pulses, we conservatively adopt a total uncertainty of 5\% for all channels in the main text (see table~\ref{tab:FRM_noise}).

\section{FRM linearity} \label{appx:linearity}

We evaluated the linearity of the flux reconstruction in FRM mode. To do so, we injected a slow flux signal through the common flux bias line while the fast sawtooth flux-ramp signal was active. The injected signal was a symmetric \qty{1}{\Hz} triangle wave with a peak-to-peak amplitude of approximately $3$--$4\,\Phi_0$. This slow modulation acts as a quasi-static flux offset relative to the ramp frequency and allows us to probe the linearity of the demodulated response over multiple flux quanta. Figure~\ref{fig:HEMT_only_linearity} shows the FRM demodulated flux as a function of the injected flux for channel~1, along with the residuals relative to an ideal linear response. The injected flux is calculated by rescaling the voltage bias to match the demodulated flux. The measured response is highly linear across the full range of injected flux. The residual nonlinearity is on the order of $1\,\mathrm{m}\Phi_0$, comparable to the performance reported in~\cite{mates_flux-ramp_2012}.

\section{JTWPA IP3 point measurement} \label{appx:TWPA_IP3}

\begin{figure}[htb]
    \centering
    \includegraphics[width=\columnwidth]{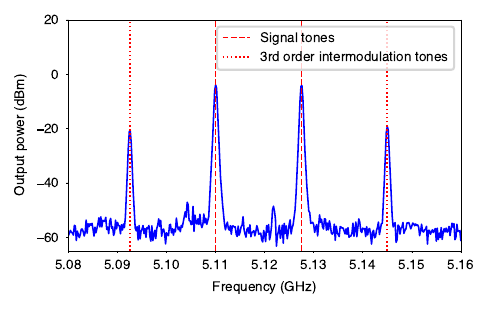}
    \caption{Spectrum of the output signal from the JTWPA and HEMT amplifier chain. The signal tones and third-order intermodulation products are marked with dotted lines.}
    \label{fig:IP3}
\end{figure}

To determine the third-order intercept point (IP3) of the JTWPA, we follow the standard two-tone measurement procedure described in~\cite{analog_devices_ip3_2013}. Two continuous-wave signal tones of equal power are injected at frequencies $f_1$ and $f_2$. Nonlinearities in the amplifier generate third-order intermodulation products at frequencies $2f_1 - f_2$ and $2f_2 - f_1$, whose powers are measured at the output of the amplification chain with a spectrum analyzer (see figure~\ref{fig:IP3}). The IIP3 is calculated as
\begin{equation}
    \text{IIP3} = P_{\text{out}} - \frac{P_{\text{out}} - P_{\text{IM}}}{2} - G,
\end{equation}
where $P_{\text{out}}$ is the measured output power of the fundamental tones, $P_{\text{IM}}$ is the measured power of the third-order intermodulation products, and $G$ is the total gain of the amplifier chain, including the JTWPA, HEMT, and room temperature amplifiers. The gain $G$ is calibrated using the attenuation of the readout line determined in~\cref{results:power_calib}.

The measurement is repeated for several input powers below the \qty{1}{\dB} compression point ($\approx-85$ to \qty{-90}{dBm}), and the resulting IIP3 values are averaged. To verify consistency across the operating band, the procedure is also performed at multiple tone frequencies within the 5--\qty{5.5}{\GHz} range. The measured IIP3 values range between \qty{-80}{dBm} and \qty{-86}{dBm}. The average IIP3 value is \mbox{\qty{-82.5}{dBm}}.

\begin{acknowledgments}
We are grateful to Kate Azar, Gabriel Cutter, and Yijie Wang for their assistance with device photography. We also thank Dr. John A. B. Mates and Dr. Manuel E. Garcia Redondo for insightful discussions. This work was supported by DOE QuantISED award DE-SC0020181, NSF under Grant PHY-2209585, and the Undersecretary of Defense for Research and Engineering under Air Force Contract No. FA8702-15-D-0001. The views and conclusions contained herein are those of the authors and should not be interpreted as necessarily representing the official policies or endorsements, either expressed or implied, of the U.S. Government or the Undersecretary of Defense for Research and Engineering.

The \ricochet project received funding from the European Research Council (ERC) under the European Union’s Horizon2020 research and innovation program under the Grant Agreement ERC-StGCENNS 803079, the French National Research Agency (ANR) within the project ANR-20-CE31-0006, the project ANR-22-EXES-0001, the LabEx Lyon Institute of Origins (ANR-10-LABX-0066) of the Université de Lyon, within the Plan France2030, Natural Sciences and Engineering Research Council of Canada (NSERC), the Canada First Excellence Research Fund, and the Arthur B. McDonald Institute (Canada). This work is also partly supported within the State Project “Science” by the Ministry of Science and Higher Education of the Russian Federation (075-15-2024-541).

\end{acknowledgments}

\bibliography{references}
\bibliographystyle{apsrev4-2}

\end{document}